\documentclass[a4paper,12pt]{article}

\usepackage{graphics}

\usepackage{amssymb}

\usepackage{natbib}
\usepackage{subcaption}
\usepackage{bstnotations}

\usepackage{mathtools}

\usepackage[bookmarksopen=true]{hyperref}

\usepackage{authblk}

\title{Using sparse polynomial chaos expansions for the global sensitivity analysis of groundwater lifetime expectancy in a multi-layered hydrogeological model}
\author[1]{G. Deman}
\author[2]{K. Konakli\thanks{Corresponding author, e-mail: konakli@ibk.baug.ethz.ch}}
\author[2]{B. Sudret}
\author[1]{J. Kerrou}
\author[1]{P. Perrochet}
\author[3]{H. Benabderrahmane}
\affil[1]{The Centre for Hydrogeology \& Geothermics (CHYN), University of Neuch\^atel, Rue Emile Argand 11, CH-2000 Neuch\^atel, Switzerland}
\affil[2]{ETH Z\"{u}rich, Institute of Structural Engineering,
Chair of Risk, Safety \& Uncertainty Quantification, Stefano-Franscini-Platz 5, CH-8093 Z\"{u}rich, Switzerland
}
\affil[3]{Andra, 1-7 rue Jean Monnet, 92298 Ch\^{a}tenay-Malabry Cedex, France}
\date{}

\graphicspath{{./},{../Figures/}}

\begin{document}

\maketitle

\begin{abstract}

  The study makes use of polynomial chaos expansions to compute Sobol'
  indices within the frame of a global sensitivity analysis of
  hydro-dispersive parameters in a simplified vertical cross-section of
  a segment of the subsurface of the Paris Basin. Applying conservative
  ranges, the uncertainty in 78 input variables is propagated upon the
  mean lifetime expectancy of water molecules departing from a specific
  location within a highly confining layer situated in the middle of the
  model domain. Lifetime expectancy is a hydrogeological performance
  measure pertinent to safety analysis with respect to subsurface
  contaminants, such as radionuclides. The sensitivity analysis
  indicates that the variability in the mean lifetime expectancy can be
  sufficiently explained by the uncertainty in the petrofacies, \ie the
  sets of porosity and hydraulic conductivity, of only a few layers of
  the model. The obtained results provide guidance regarding the
  uncertainty modeling in future investigations employing detailed
  numerical models of the subsurface of the Paris Basin. Moreover, the
  study demonstrates the high efficiency of sparse polynomial chaos
  expansions in computing Sobol' indices for high-dimensional models. \\
  
  {\bf Keywords} -- global sensitivity analysis -- polynomial chaos
  expansions -- groundwater flow -- lifetime expectancy -- deep
  geological storage

\end{abstract}

\section{Introduction}
\label{sec:01}

With the improvement of computing power, numerical modeling has become a popular tool for understanding and predicting various kinds of subsurface processes addressed in the fields of geology and hydrogeology. However, the incomplete/imprecise knowledge of the underground system frequently compels the modeller to make a number of approximations and assumptions with regard to the geometry of geological structures, the presence of discontinuities and/or the spatial distribution of hydro-dispersive parameters in their models \cite{Renard2007}. These uncertainties can possibly lead to large variabilities in the predictive modeling of subsurface processes and thus, it becomes of major importance to account for the aforementioned assumptions in the frame of uncertainty and sensitivity analyses. Uncertainty analysis (UA) aims at quantifying the variability of a given response of interest as a function of uncertain input factors, whereas sensitivity analysis (SA) has the purpose to identify the input factors responsible for this variability. Hence, SA determines the key variables to be described in further detail in order to reduce the uncertainty in the predictions of a model.

Methods of SA are typically classified in two categories: local SA and global SA methods. The former investigate effects of variations of the input factors in the vicinity of nominal values, whereas the latter aim at quantifying the output uncertainty due to variations in the input factors in their entire domain. Among several global SA methods proposed in the literature, of interest herein is SA with \emph{Sobol' sensitivity indices}, which belongs to the broader class of variance-based methods \cite{Saltelli2008}. These methods rely upon the decomposition of the response variance as a sum of contributions of each input factor or combinations thereof and do not assume any kind of linearity or monotonicity of the model. We note that the Fourier amplitude sensitivity test (FAST) \cite{Cukier1978, Saltelli1999} indices enter this class as well.

Various methods have been investigated for computing the Sobol' indices that were first defined in \cite{Sobol1993}, see \eg \cite{Archer1997, Sobol2001, Saltelli2002, Sobol2005, Saltelli2010}. In these papers, Monte Carlo simulation is used as a tool to estimate these sensitivity indices. This has revealed extremely costly, although more efficient estimators have been recently proposed \cite{Sobol2007, Janon2013}. In the recent years, new approaches using surrogate models have been introduced in the field of global SA \cite{Oakley2004, Marrel2009, Storlie2009, zuniga2013}. A popular method to compute the Sobol' indices, originally introduced by \cite{SudretRESS2008b}, is by post-processing the coefficients of a polynomial chaos expansion (PCE) meta-model of the response quantity of interest. PCE constitutes an efficient UA method in which the key concept is to expand the model response onto a basis made of orthogonal polynomials in the input variables. Once a PCE representation is available, the Sobol' indices can be calculated analytically with elementary operations at almost no additional computational cost. Sparse PCE make the approach even more efficient, as shown in \cite{BlatmanRESS2010}.

In the frame of the stochastic modeling of subsurface flow and mass transport, PCE meta-models have proven to be comprehensive and robust tools for performing SA at low computational cost. As an example, applying a PCE-based global SA upon a fine-grid numerical model of flow and mass transport in a heterogeneous porous medium, \citet{Fajraoui2011} and \citet{younes2013} established the transient effect of uncertain flow boundary conditions, hydraulic conductivities and dispersivities on solute concentrations at given observation points. \citet{Sochala2013} propagated uncertain soil parameters upon three different physical models of subsurface unsaturated flow. Their study proved the higher efficiency of PCE meta-models, in comparison to a classical Monte-Carlo method, for representing the variability of the output quantity at low computational cost. In the frame of radionuclide transport simulation in aquifers, \citet{Ciriello2013} analyzed the statistical moments of the peak solute concentration measured at a specific location, as a function of the conductivity field, the dispersivity coefficients and the partition coefficients associated to the heterogeneous media. The comparison of the Sobol' indices obtained for various degrees of PCE meta-models showed that low-degree models can yield reliable indices while considerably reducing the computational burden. \citet{formaggia2013} used PCE-based sensitivity indices to investigate effects of uncertainty in hydrogeological variables on the evolution of a basin-scale sedimentation process. However, the various aforementioned contributions consider simplified models for the description of subsurface flow and mass transport. A detailed site characterization model was employed by \citet{laloy2013efficient}, but global SA was confined to flow processes.

In the scope of the deep geological storage of radioactive wastes, ANDRA (French National Radioactive Waste Management Agency) has conducted several studies to assess the potentiality of a clay-rich layer for establishing a mid to long-lived radioactive waste disposal in the subsurface of the Paris Basin. The thick impermeable layer from Callovo-Oxfordian (COX) age has been extensively studied \citep{Delay2006,Distinguin2007,Enssle2011} together with the two major limestone aquifers, in place of the Dogger and the Oxfordian sequences \citep{Brigaud2010,Linard2011,Landrein2013}, encompassing the claystone formation. A recent study \citep{Deman2015} used a high-resolution integrated Meuse/Haute-Marne hydrogeological model \citep{ANDRA2012} to compute the average time for water molecules departing from a given area in the COX to reach the limits of the domain where the numerical model is defined. SA over hydro-dispersive parameters in 14 hydrogeological layers proved that the Dogger and Oxfordian limestone sequences have a large influence on the residence time of groundwater. Indeed, advection processes occurring in permeable layers strongly influence the water transit in the subsurface of the Paris Basin, in contrast to the slow-motion diffusive processes taking place in impermeable rocks.

However, the analysis of the effect of uncertainties related to other advective-dispersive parameters, such as boundary conditions, orientations and anisotropies of hydraulic conductivity tensors or magnitudes of dispersion parameters, represents a great effort that cannot be carried out with the integrated model at reasonable computational costs. Addressing the issue of performing UA with the use of high-resolution numerical models of geological reservoirs, Castellini and co-workers \citep{Castellini2003} established that numerical models built at the coarse scale, but covering a reasonable number of geological and geostatistical features, can be particularly informative in capturing the main subsurface processes at low computational costs.

The present study introduces a vertical two-dimensional multi-layered hydrogeological model representing a simplification of the underground media of the Paris Basin in the vicinity of the site of Bure and does not integrate the complex geometry of the layers, neither does it include the numerous discontinuities or heterogeneities observed in the field. It must be emphasized that this simplified model is not aimed at site characterization, but at evaluation of hydrogeological performance through SA for calibration purposes.

The main objective of the present work is to assess the effect of multiple advective-dispersive parameters on the mean lifetime expectancy (MLE) of water molecules departing from a target zone in the central layer. The MLE corresponds to the average time required for a given solute at a specific location to reach any outlet of the model domain and is a critical quantity in safety analysis dealing with subsurface contaminants such as radionuclides. This work represents a substantial complement to the study by \citet{Deman2015} by encompassing a large scope of uncertain factors, which cannot be assessed using the integrated model due to the computational burden. Conservative uncertainty ranges are defined for the input factors analyzed in the frame of a SA relying on the estimation of PCE-based Sobol' indices. The sparse PCE approach was chosen because of its ability to tackle high-dimensional problems with great efficiency. The study provides recommendations for future investigations employing the high-resolution integrated Meuse/Haute-Marne hydrogeological numerical model of the Paris Basin; in particular, it identifies the sets of parameters that can be fixed to their nominal values without significantly affecting the MLE variability as well as the sets of parameters to be described in further detail.

The paper is organized as follows: The subsequent Section 2 provides a comprehensive description of the considered hydrogeological model, while Section 3 presents the concepts of SA with Sobol' indices and the computation of those indices using PCE. Section 4 includes the results of UA and SA of the model, along with interpretations accounting for the underlying physics. Finally, the conclusive Section 5 summarizes the study, highlighting the main findings of the above analyses, and provides recommendations for future investigations.

\section{The numerical model}
\label{sec:02}

\subsection{Geometry and finite element mesh}

\begin{figure}[!ht]
\begin{center}
\includegraphics[width=0.98\textwidth]{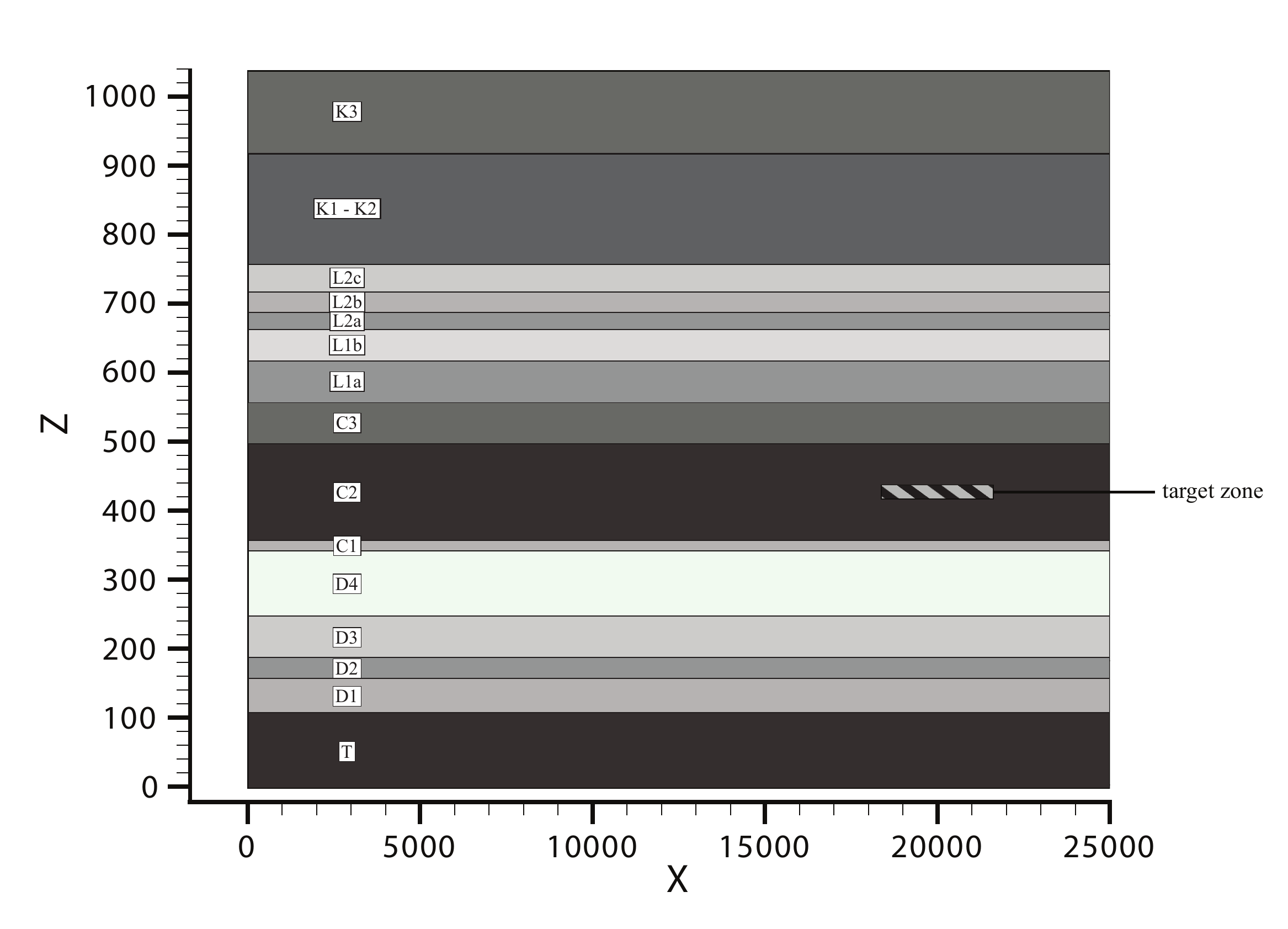}
\caption{Geometry and geological layers with the localization of the target zone (vertical exaggeration: 20).}
\label{fig:UPSILONhom}
\end{center}
\end{figure}

Originally inspired by the COUPLEX numerical model from \citet{Bourgeat2004}, the present model stands as a vertical two-dimensional ($x$-$z$) cross-section of 25,000 $\times$ 1,040 meters representing a segment of the Paris Basin subsurface. The mesh is discretized into 5 $\times$ 5 meters square elements for a total of 1,040,000 elements. In order to subdivide the domain into entities related to geological formations, the main features of the subsurface were extracted from the lithostratigraphic log of the deep EST433 borehole \citep{Landrein2013} in the vicinity of the experimental site of Bure (Haute-Marne, France). Therefore, the model consists of 15 hydrogeological layers characterized by tabular geometries, uniform thicknesses and homogeneous parameters. Figure \ref{fig:UPSILONhom} summarizes the geometry of the model and gives an overview on the succession and thicknesses of layers.

The bottom layer stands as a 110~m thick low-permeability layer attributed to the Toarcian marl formation (T). Overlying the latter, the succession of carbonate formations from the Dogger sequence is subdivided into 5 layers of which the total thickness attains 250~m in the spatial domain. The sequence encompasses the Bajocian (D1) and Bathonian limestones (D3 and D4) representing the main aquifer formations of the Dogger, a clastic dominated interval (\emph{"Marnes de Longwy"}, D2) separating the two. The Dogger sequence is topped with a thin oolithic limestone from Lower Callovian (\emph{"Dalle Nacr\'{e}e"}, C1), implemented as a 15~m thick layer in the model. The latter marks the transition with the thick, highly impermeable, claystone formation of Callovo-Oxfordian age (C2) of which the thickness reaches 150~m in the model. In the numerical simulations, a target zone (TZ) located in the middle of layer C2 (Figure \ref{fig:UPSILONhom}) represents the location for the computation of the output quantity of interest.

The low-permeability COX layer is overlaid by a limestone sequence of the Oxfordian age. The latter is incorporated as a 260~m thick formation subdivided into 6 hydrogeological entities. A relatively confining layer from the Upper Argovian (C3ab) rests directly on the COX and is followed by permeable formations of the Rauracian-Sequanian sequence (L1a to L2c). A thick interval of marls and argillaceous limestones from Kimmeridgian age (K1-K2) covers the whole and is implemented as a 160~m thick low-permeability layer. The top layer is a 120~m thick confining formation attributed to the Tithonian (K3). The latter outcrops in the vicinity of Bure.

\subsection{Governing equations and model outputs}    	
\label{sec:Equations}

In the numerical simulations the flow is governed by the steady-state equation
\begin{equation}\label{eq:Flux}
\nabla \cdot \mathbf{q} = 0,
\end{equation}
where $\mathbf{q} = - \mathbf{K} \; \nabla H$, is the Darcian flux vector [L T$^{-1}$], \textbf{K} is the tensor of hydraulic conductivity [L T$^{-1}$] and $H$ is the hydraulic head [L]. The anisotropy $A_K$ in the components of the tensor of hydraulic conductivity is defined as the ratio between the hydraulic conductivities in the two principal directions: $A_K = K_z / K_x$.

Here, it is assumed that \textbf{K} has orthotropic properties. Considering a hydraulic conductivity tensor $\mathbf{K}_p$ of which the components are mapped into the Cartesian system and given along their principal direction, $X_p$, the tensor \textbf{K} in the global Cartesian space is retrieved by means of the rotation matrix \textbf{R} with the expression
\begin{equation}\label{eq:Ktensor}
\mathbf{K} = \mathbf{R}^T \; \mathbf{K}_p \; \mathbf{R}.
\end{equation}
For the two-dimensional problem considered, the rotation matrix \textbf{R} is defined in terms of the Euler angle $\theta$ (in degree) as
\begin{equation}\label{eq:RotationMat}
\mathbf{R} = \begin{pmatrix}
\cos \theta & \sin \theta \\
- \sin \theta & \cos \theta
\end{pmatrix}.
\end{equation}

In the present study, steady-state flow simulations are carried out together with the computation of the lifetime expectancy probability density function (PDF) at any point $x$  in the domain.
Under stationary conditions (\ie steady-state flow), the lifetime expectancy PDF addresses the probability distribution of the time required for a solute, taken at any position $x$, to leave the domain. In its formulation, the lifetime expectancy PDF assimilates the forward advective-diffusive transport equation (ADE) to the Fokker-Planck (forward Kolmogorov) equation measuring the random motion of solute particles \citep{Uffink1989}. For more details on the computation of the lifetime expectancy PDF, the reader is referred to \citet{Cornaton2006,Cornaton2006a} and \citet{Kazemi2006}.

Based on the ADE, the lifetime expectancy PDF is computed using the backward transport equation requiring reversed flow directions ($\textbf{q} := -\textbf{q}$) as well as adjusted downstream boundary conditions. The lifetime expectancy PDF $g_E(x,t)$ at any point $x$ in the domain is then governed by
\begin{equation}\label{eq:LE}
\phi \frac{\partial g_E}{\partial t} = \nabla \cdot (\mathbf{q} \; g_E + \mathbf{D} \; \nabla g_E),
\end{equation}
where $\phi$ is the effective porosity [-] and where \textbf{D} is the dispersion tensor
\begin{equation}\label{eq:Disp}
\phi \mathbf{D} = (\alpha_L - \alpha_T) \frac{\mathbf{q} \otimes \mathbf{q}}{|| \mathbf{q} ||} \; + \; \alpha_T  \; || \mathbf{q} || \; \mathbf{I} \; + \; \phi \; D_m \; \mathbf{I},
\end{equation}
where \textbf{I} is the identity matrix, $D_m$ is the coefficient of molecular diffusion [L$^2$ T$^{-1}$], $\alpha_L$ and $\alpha_T$ are the longitudinal and transverse components of the macro-dispersion tensor [L] respectively. In the present study, the anisotropy in the macro-dispersion tensor is determined with the coefficient $A_\alpha = \alpha_T / \alpha_L$.

The straightforward computation of the first moment of the lifetime expectancy PDF is the so-called \emph{mean lifetime expectancy} (MLE) $E(x,z)$ [T] at any position $x$, governed by
\begin{equation}\label{eq:MLE}
- \nabla \cdot (\mathbf{q} \; E + \mathbf{D} \; \nabla E) = \phi,
\end{equation}
where it can be seen that the porosity $\phi$ [-] acts as the sink term in the aging process.

The TZ comprises a set of 1,947 nodes in layer C2, covered by a rectangle of which the lateral and vertical extensions are $x$ = [18,440; 21,680], $z$ = [425; 435] (Figure \ref{fig:UPSILONhom}). In the present study, the arithmetic mean of $E(x,z)$ calculated at each of these 1,947 nodes stands for the output response of interest and is used in the subsequent analysis. It can be seen as the average time for a solute originating from the TZ to reach  any outlet of the domain of the model.

The finite element simulator \emph{GroundWater} \citep{Cornaton2007} was employed to solve Eq.~(\ref{eq:Flux})-(\ref{eq:MLE}) using the finite element technique. A single run of steady-state flow and MLE computation takes about 120 seconds using a parallel solver with 6 CPU.

The reader should note that the use of a 2D vertical model to solve for the hydro-dispersive processes cannot capture correctly the real behavior of the Paris Basin subsurface because, apart from being a simplified model, it omits the lateral flow and dispersion along the third dimension. This has the effect of underestimating the magnitude of the modeled processes \cite{Kerrou2010}. It is however assumed that this bias is equivalent for all the layers considered and thus, the interpretation of the SA results obtained with the 2D cross-section may be generalized to a synthetic 3D case employing the same settings.

\subsection{Flow boundary conditions}
\label{sec:FlowBCs}

\begin{figure}[!ht]
\begin{center}
\includegraphics[width=0.98\textwidth]{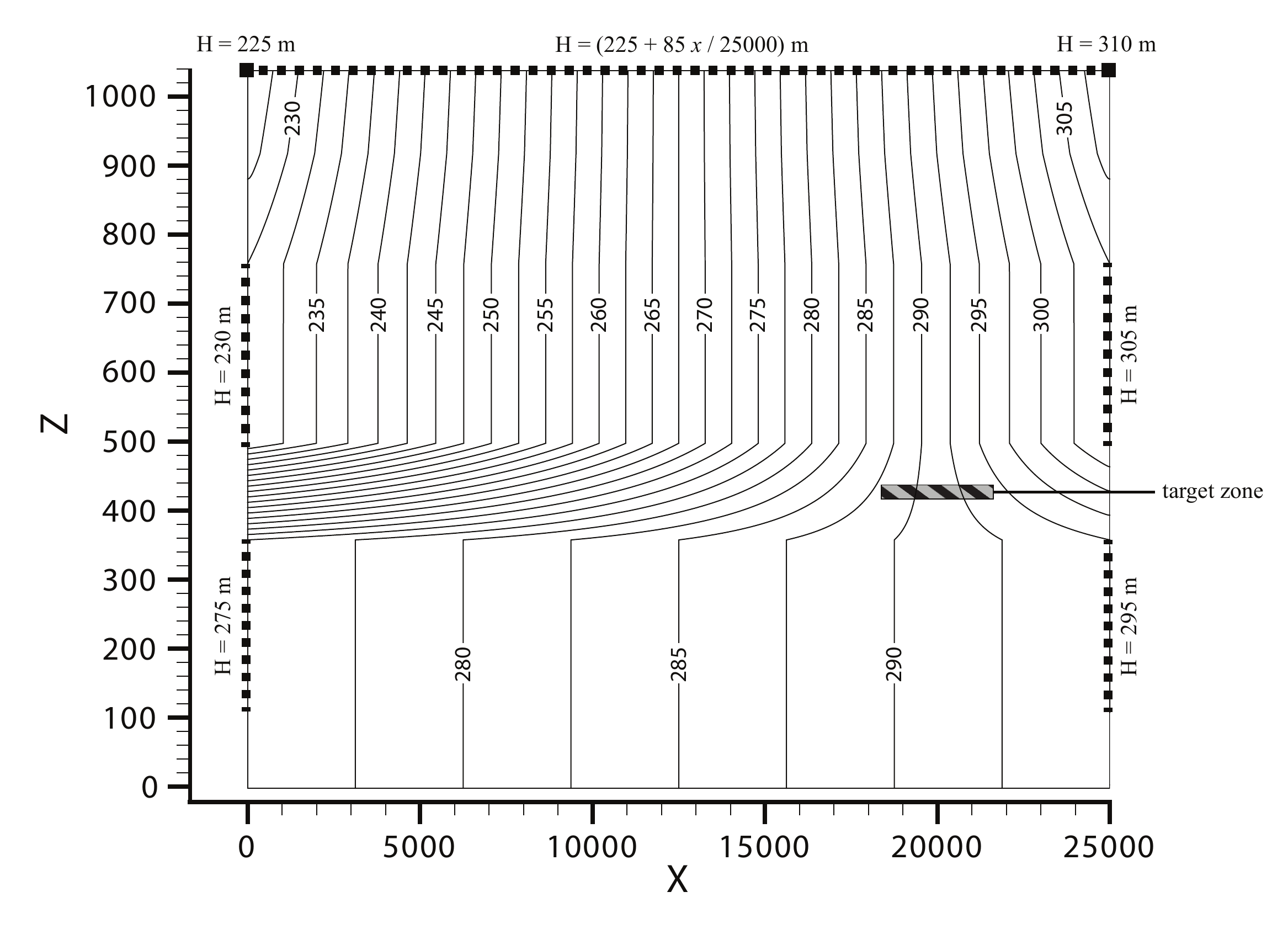}
\caption{Flow boundary conditions and head contours (vertical exaggeration: 20).}
\label{fig:isopiezes}
\end{center}
\end{figure}

The fully saturated model considers stationary flow conditions in a confined aquifer which are implemented as Dirichlet type flow boundary conditions (BCs). These flow BCs are imposed on nodes located at the top of the model domain as well as at both lateral limits of the two limestone sequences (Figure \ref{fig:isopiezes}).

Regional piezometric maps based on field measures \citep{Linard2011} were used to constrain the hydraulic gradients in both carbonate sequences.
The flow BCs imposed on the lateral boundaries of the two limestone sequences derive from a 25~km transect starting from the Gondrecourt trough and extending in a North-West direction, the main regional flow direction. The hydraulic gradient set on top of the domain corresponds to the average topographic gradient of the region covered by the transect.

Under these conditions, the general groundwater flow direction is oriented from right to left. The proportions of the total outflowing rates are approximately 2\%, 60\% and 38\% for the top of the domain, the Oxfordian and the Dogger discharge boundaries, respectively. In layer C2, the groundwater flows downward in the very right part of the domain and then upward in the remainder, with a hydraulic gradient inversion in the vicinity of the TZ (see Figure \ref{fig:isopiezes}). As a summary, the flow BCs are gathered in Table \ref{tab:BCs}.

\begin{table}[!ht]
\caption {Flow boundary conditions.}
\label{tab:BCs}
\centering
\begin{tabular}{l l l}
\hline
Boundary & Position & Hydraulic head \\
\hline
right Oxfordian  & $x$ = 25000, $z$ = [500, 760] & $H$ = 305 m \\
left Oxfordian   & $x$ = 0, $z$ = [500, 760]     & $H$ = 230 m \\
right Dogger     & $x$ = 25000, $z$ = [110, 360] & $H$ = 295 m \\
left Dogger      & $x$ = 0, $z$ = [110, 360]     & $H$ = 275 m \\
top of domain & $x$ = [0, 25000], $z$ = 1040  & $H = 225 + 85 x / 25000$ \\
\hline
elsewhere & & \emph{no flow} \\
\hline
\end{tabular}
\end{table}

To account for uncertainties in the flow BCs, the hydraulic gradients in the two limestone sequences and on the top of the domain are considered as uncertain input factors included in the following SA (Section \ref{sec:04}). A change in the hydraulic gradients may shift the position of the vertical groundwater flux inversion in layer C2, and thus the MLE calculated at the TZ.

\subsection{Hydraulic conductivity and porosity values}

Many studies have undertaken the inventory of hydraulic conductivity ($K$) and porosity ($\phi$) values in the various geological formations of the Paris Basin. For a large number of wells and boreholes within a wide area around the experimental site of Bure, laboratory and field measurements were conducted to provide $\{K,\phi\}$ datasets for the two limestone sequences \citep{Brigaud2010,Linard2011,Fourre2011,Delay2004,Delay2007}. 

However, very few $\{K,\phi\}$ datasets are available for the four low-permeability formations implemented in the present model (\ie K3, K1-K2, C2 and T). Hence, data extracted from the literature \citep{Cosenza2002,Delay2007a,Delay2006,Enssle2011,Mazurek2011,Vinsot2011}, and employed in previous studies \citep{Contoux2013,Hoyos2012,Goncalves2004,Goncalves2004a}, were used to define the uncertainty ranges for the $\{K,\phi\}$ sets in these layers.

In the geological formations of the Oxfordian and Dogger sequences, large variabilities of the $\{K,\phi\}$ couples are noticed with the presence of dependencies (\eg low $K$ and low $\phi$ values are correlated).
However, in order to simplify the conceptual approach in a first stage, a perfect dependence between $\log_{10}(K)$ and $\phi$ is defined by making use of a mathematical function approximating the relationship between these two in each layer. In the sequel, both parameters are referred to as a whole under the name \emph{petrofacies}. This approach reduces the computational burden of the subsequent SA by avoiding the use of correlation functions between the two uncertain factors. The hydraulic conductivity values deriving from the $\{K,\phi\}$ distributions in each layer are attributed to the longitudinal component of the hydraulic conductivity tensor, $K_x$.
For each layer, the estimated value of $K_x$ is retrieved through a relationship: $\log_{10}(K_x)=f(\phi)$ \citep{Deman2015}. 

Although no explicit information is available on the following, the geological formations are believed to feature anisotropic hydraulic conductivity tensors \textbf{K}, \ie anisotropy in the two principal components of the tensor ($K_x$ and $K_z$) defined as the ratio $A_K = K_z / K_x$. In the nominal case, $A_K$ = 0.1 is assumed for every layer in the model.

Preferential flow directions are supposedly taking place within each individual layer. For each layer, the Euler angle $\theta$ defines the orientation of the hydraulic conductivity tensor $\mathbf{K}_p$ in the Cartesian space (see Eq.~(\ref{eq:Ktensor})-(\ref{eq:RotationMat})). In the nominal case, $\theta = 0$ degree is assumed in every layer, which corresponds to the two principal components of the hydraulic conductivity tensors $\mathbf{K}_p$ being oriented along the $x$ and $z$ axes. The orientation of the groundwater flux $\mathbf{q}$ in the model is principally due to the static hydraulic gradients $\nabla H$ resulting from flow BCs implemented on the  edges of the domain. Note however that the Euler angle $\theta$ may locally change the orientation of $\mathbf{q}$ in a given layer and thus, drive the groundwater into adjacent layers where magnitudes might be different. This phenomenon may have a significant effect on the MLE calculated from the TZ, which is explored in Section \ref{sec:04}.

Table \ref{tab:NominalValues} summarizes the nominal values for $\phi$ and the corresponding $K_x$ in each of the 15 hydrogeological layers comprised in the model. The values for $\phi$ correspond to the mean value (or the median value of the CDF) of the distribution in each layer, whereas the values for $K_x$ derive from approximation functions.
As a reminder, the present study assumes homogeneous parameters in every layer. Although this feature is unrealistic, it is recalled that the purpose of this study is to bring insights into the global effect of equivalent advective-dispersive parameters of the multi-layered hydrogeological model and to provide recommendations for a similar application on a high-resolution integrated hydrogeological model of the Paris Basin.

\begin{table}[!ht]
\caption{Nominal values for the porosity ($\phi$) and the longitudinal hydraulic conductivity ($K_x$) in the 15 hydrogeological layers.}
\label{tab:NominalValues}
\centering
\begin{tabular}{c c c}
\hline
Layer &$K_x$ [m/s]&$\phi$ [-]\\
\hline
K3 & 9.01E$-09$ & 0.0100\\
K1-K2 & 4.53E$-09$ & 0.1150\\
L2c & 1.10E$-06$ & 0.1389\\
L2b & 3.46E$-07$ & 0.1110\\
L2a & 1.62E$-07$ & 0.1139\\
L1b & 1.49E$-05$ & 0.1604\\
L1a & 1.17E$-06$ & 0.1549\\
C3ab & 4.59E$-08$ & 0.0984\\
C2 & 1.99E$-13$ & 0.1580\\
C1 & 1.89E$-06$ & 0.0470\\
D4 & 1.65E$-05$ & 0.0905\\
D3 & 1.76E$-06$ & 0.1016\\
D2 & 2.62E$-07$ & 0.0623\\
D1 & 3.23E$-06$ & 0.0688\\
T & 1.95E$-12$ & 0.0810\\
\hline
\end{tabular}
\end{table}


\subsection{Dispersion parameters}    	

The mean lifetime expectancy formulation (Eq.~(\ref{eq:MLE})) is an advective-dispersive solute transport equation, where the longitudinal and transverse components of the macro-dispersion tensor ($\alpha_L$ and $\alpha_T$ respectively) control the particles dispersion. These two uncertain factors depend particularly on the rock type, on the tortuosity of the porous media and also, on the scale considered. Homogeneous values of $\alpha_L$ and $\alpha_T$ are set within each layer, with the values $\alpha_L$ = 15~m and $A_{\alpha} = \alpha_T / \alpha_L$ = 0.1 considered in the  entire domain of the model in the nominal case.

As mentioned previously, no decay or adsorption effects are accounted in the computation of the MLE. 
The coefficient of molecular diffusion corresponds to the theoretical self-diffusion coefficient for the water molecule, $D_m = 2.3 \; 10^{-9}$ m$^2$/s.

\section{Polynomial chaos expansions for sensitivity analysis }
\label{sec:03}
Let us denote by $\cm$ the computational model describing the behavior of the considered physical system. Let $\ve X=\{X_1 \enum X_M\}$ denote the $M$-dimensional random input vector with joint PDF $\pdf$ and marginal PDFs $f_{X_i}(x_i)$, $i=1 \enum M$. Due to the input uncertainties represented by $\ve X$,  the model response of interest becomes random. The computational model is thus considered as the map
\begin{equation}
\label{eq:model}
\ve X \in \cd_{\ve X} \subset \Rr^M\longmapsto Y=\cm (\ve X) \in \Rr,
\end{equation}
where $\cd_{\ve X}$ is the support of $\ve X$. In the description of the theoretical framework hereafter, we assume that the components of $\ve X$ are \emph{independent}, which is the case for the model in the present study.

As explained in the Introduction, the aim of global SA is to identity random input variables and combinations thereof with significant contributions to the variability of $Y$ as described by its variance. A concise description of the employed method of PCE-based Sobol' sensitivity indices is given in the following; for further details on the topic, the reader is referred to \cite{SudretRESS2008b} and \cite{BlatmanRESS2010}. The extension to the case of mutually dependent random variables is  presented in \cite{LiRabitz2010, Kucherenko2012, Mara2012variance}.

\subsection{Sobol' indices}

Assuming that the function $\cm(\ve X)$ is square-integrable with respect to the probability measure associated with $\pdf$, the Sobol' decomposition of $Y=\cm(\ve X)$ in summands of increasing dimension is given by \cite{Sobol1993}
\begin{equation}
\label{eq:Sobol_long}
\cm(\ve X) = \cm_0+\sum_{i=1}^M\cm_i(X_i)+\sum_{1\leq i< j\leq M}\cm_{ij}(X_i,X_j)+ \ldots+M_{12\ldots M}(\ve X)
\end{equation}
or equivalently, by
\begin{equation}
\label{eq:Sobol_short}
\cm(\ve X) = \cm_0+\sum_{\ve u\neq\emptyset}\cm_{\ve u}(\ve X_{\ve u}),
\end{equation}
where $\cm_0$ is the mean value of $Y$, $\ve u=\{i_1 \enum i_s\}\subset\{1 \enum M\}$ are index sets and $\ve X_{\ve u}$ denotes a subvector of $\ve X$ containing only those components of which the indices belong to $\ve u$. The number of summands in the above equations is  $2^M-1$.

The Sobol' decomposition is unique under the condition
\begin{equation}
\label{eq:Sobol_uniq}
\int_{\cd_{X_k}}\cm_{\ve u}(\ve x_{\ve u})f_{X_k}(x_k)dx_k=0, \hspace{5mm} \mathrm{if} \hspace{1mm} k\in \ve u,
\end{equation}
where $\cd_{X_k}$ and $f_{X_k}(x_k)$ respectively denote the support and marginal PDF of $X_k$. Eq.~(\ref{eq:Sobol_uniq}) leads to the orthogonality property
\begin{equation}
\label{eq:Sobol_orthog}
\Espe{}{\cm_{\ve u}(\ve X_{\ve u})\cm_{\ve v}(\ve X_{\ve v})}=0,\hspace{5mm} \mathrm{if} \hspace{1mm} \ve u\neq \ve v.
\end{equation}
The uniqueness and orthogonality properties allow decomposition of the variance $D$ of $Y$ as
\begin{equation}
\label{eq:var}
D=\Var{\cm(\ve X)}=\sum_{\ve u\neq\emptyset}D_{\ve u},
\end{equation}
where $D_{\ve u}$ denotes the partial variance
\begin{equation}
\label{eq:partial_var}
D_{\ve u}=\Var{\cm_{\ve u}(\ve X_{\ve u})}=\Espe{}{\cm^2_{\ve u}(\ve X_{\ve u})}.
\end{equation}

The Sobol' index $S_{\ve u}$ is defined as
\begin{equation}
\label{eq:Sobol_index}
S_{\ve u}=D_{\ve u}/D,
\end{equation}
and describes the amount of the total variance that is due to the interaction between the uncertain input parameters comprising $\ve X_{\ve u}$. By definition, $\sum_{\ve u\neq\emptyset}S_{\ve u}=1$.
First-order indices, $S_i$, describe the influence of each parameter $X_i$ considered separately, also called \textit{main effects}. Second-order indices, $S_{ij}$, describe the influence from the interaction between the parameters $\{X_i,X_j\}$. Higher-order indices describe influences from interactions between larger sets of parameters.

The total sensitivity indices, $S^T_i$, represent the \textit{total effect} of an input parameter $X_i$, accounting for its main effect and all interactions with other parameters. They are derived from the sum of all partial sensitivity indices $S_{\ve u}$ that involve parameter $X_i$, \ie
\begin{equation}
\label{eq:Sobol_total_index}
S^T_i=\sum_{\ci_i}D_{\ve u}/D,\hspace{5mm} \ci_i=\{\ve u\supset i\}.
\end{equation}
It follows that $S^T_i=1-S_{\sim i}$, where $S_{\sim i}$ is the sum of all $S_{\ve u}$ with $\ve u$ not including $i$.

Evaluation of the Sobol' indices by Monte Carlo simulation is based on a recursive relationship that requires computation of $2^M$ Monte Carlo integrals involving $\cm(\ve X)$ in order to obtain the full set. Clearly, this is not affordable when the computational model is a time-consuming algorithmic sequence. In typical applications, first-order and total indices are computed. We note that the first-order indices are equivalent to the first-order indices obtained by the FAST method, which may provide a more efficient computation.  It will be shown next that when a PCE of the quantity of interest is available, the \emph{full} set of Sobol' indices can be obtained analytically at almost no additional computational cost.

\subsection{Polynomial chaos expansions}

\subsubsection{Computation of polynomial chaos expansions}
\label{sec:PCE}

A PCE approximation of $Y=\cm (\ve X)$ in Eq.~(\ref{eq:model}) has the form \cite{Xiu2002}
\begin{equation}
\label{eq:PCE}
\widehat{Y}=\cm^{\rm{PCE}}(\ve X)=\sum_{\ua \in \ca}{y_{\ua}} \Psi_{\ua}(\ve {X}),
\end{equation}
where $\ca$ is a set of multi-indices $\ua=(\alpha_1 \enum \alpha_M)$, $\{\Psi_{\ua}, \ua \in \ca\}$ is a set of multivariate polynomials that are orthonormal with respect to $f_{\ve X}(\ve x)$ and $\{y_{\ua}, \ua \in \ca\}$ is the corresponding set of polynomial coefficients.

The multivariate polynomials that comprise the PCE basis are obtained by tensorization of appropriate univariate polynomials, \ie
\begin{equation}
\label{eq:mult_pol}
\Psi_{\ua}(\ve X)=\prod_{i=1}^M\psi^{(i)}_{\alpha_i}(X_i),
\end{equation}
where $\psi^{(i)}_{\alpha_i}(X_i)$ is a polynomial of degree ${\alpha_i}$ in the $i$-th input variable belonging to a family of polynomials that are orthonormal with respect to $f_{X_i}(x_i)$. For standard distributions, the associated family of orthonormal polynomials is well-known \eg a standard normal variable is associated with the family of Hermite polynomials, whereas a uniform variable over $[-1,1]$ is associated with the family of Legendre polynomials. A general case can be treated through an isoprobabilistic transformation of the input random vector $\ve X$ to a basic random vector \eg a vector with independent standard normal components or independent uniform components over $[-1,1]$. The set of multi-indices $\ca$ in Eq.~(\ref{eq:PCE}) is determined by an appropriate truncation scheme. In the present study, a hyperbolic truncation scheme is employed, which corresponds to selecting all multi-indices that satisfy
\begin{equation}
\label{eq:trunc}
\norme {\ua} q = \Bigg(\sum_{i=1}^M \alpha_i^q\Bigg)^{1/q}\leq p. 
\end{equation}
for appropriate $0< q\leq 1$ and $p \in \mathbb{N}$ \cite{BlatmanPEM2010}. When $q=1$, polynomials of maximum total degree $p$ are retained, whereas use of a lower $q$ limits the number of basis terms that include interactions between two or more variables. Optimal values of these parameters may be selected in terms of error estimates \eg by using cross validation techniques.

Once the basis has been specified, the set of coefficients $\ve y=\{y_{\ua},\hspace{1mm} \ua \in \ca\}$ may be computed by minimizing the mean-square error of the approximation over a  set of $N$ realizations of the input vector, $\ce=\{\ve x^{(1)} \enum \ve x^{(N)}\}$, called \emph{experimental design}. Efficient solution schemes are obtained by considering the regularized  problem
\begin{equation}
\label{eq:LAR}
\ve y= \underset{\ve {\upsilon}\in\Rr^{\mathrm{card}\ca}}{\mathrm{arg\hspace{1mm} min}}\hspace{1mm}\sum_{i=1}^N\bigg(\cm(\ve x^{(i)})-\sum_{\ua \in \ca}\upsilon_{\ua}\Psi_{\ua}(\ve x^{(i)})\bigg)^2+C\|\ve \upsilon\|_1^2,
\end{equation}
where $\|\ve \upsilon \|_1=\sum_{j=1}^{\rm{card}\,\ca}|\upsilon_j|$ and $C$ is a non-negative constant. A nice feature of the above regularized problem is that it provides a \emph{sparse} meta-model by disregarding insignificant terms from the set of predictors. Higher values of $C$ lead to sparser meta-models, while its optimal value is typically identified as the one leading to the minimum error estimated with \eg cross validation techniques \cite{Tibshirani:1996}. In the present application, we solve Eq.~(\ref{eq:LAR}) using the \emph{hybrid Least Angle Regression} (LAR) method  as originally proposed in \cite{BlatmanJCP2011}. Hybrid LAR employs the LAR algorithm \cite{Efron2004} to select the best set of predictors and subsequently, estimates the coefficients with standard least-squares minimization.

\subsubsection{Error estimates}
\label{sec:err}

A good measure of the accuracy of PCE is the mean-square error of the residual,  $Err_G = \Esp{\left(Y-\widehat Y \right)^2}$, called \emph{generalization error}. In practice, this could be estimated by Monte Carlo simulation using a sufficiently large  set of $n_{\mathrm{val}}$ realizations of the input vector,  $\cx_{\mathrm{val}}=\{{\ve x}_1 \enum {\ve x}_{n_{\mathrm{val}}}\}$, called \emph{validation set}. The estimate of the generalization error is given by
\begin{equation}
\label{eq:errG}
\widehat{Err}_G=\frac{1}{n_{\rm {val}}}\sum_{i=1}^{n_{\rm{val}}}\bigg(\cm(\ve x_i)-\sum_{\ua \in \ca}y_{\ua}\Psi_{\ua}(\ve x_i)\bigg)^2.
\end{equation}
The relative generalization error, $\widehat{err}_G$, is estimated by normalizing $\widehat{Err}_G$ with the empirical variance of $\cy_{\mathrm{val}}=\{\cm({\ve x}_1)\enum \cm({\ve x}_{n_{\mathrm{val}}})\}$,  denoting the set of model evaluations at the validation set.

However, PCE are typically used as surrogate models in cases when evaluating a large number of model responses is not affordable. It is thus desirable to get an estimate of $Err_G$ using only the information obtained from the experimental design. One such measure is the \emph{Leave-One-Out} (LOO) \emph{error} \cite{Allen1971}. The idea of the LOO cross-validation is to set apart one point of the experimental design, say $\ve x^{(i)}$, and use the remaining points to build the PCE, denoted $\cm^{\rm{PCE} \backslash \mathit i}$. The LOO error is obtained after alternating over all points of the experimental design, \ie
\begin{equation}
\label{eq:errLOO}
{Err}_{LOO}=\frac{1}{N}\sum_{i=1}^N (\cm(\ve x^{(i)})-\cm^{\rm{PCE} \backslash \mathit i}(\ve x^{(i)}))^2.
\end{equation}
Although the above definition outlines a computationally demanding procedure, algebraic manipulations allow evaluation of the LOO error from a \emph{single} PCE based on the full experimental design. Let us denote by $h_i$ the $i$-th diagonal term of matrix $\ve \Psi(\ve \Psi^{\rm{T}}\ve \Psi)^{-1}\ve \Psi^{\rm{T}}$, where $\ve \Psi=\{\ve \Psi_{ij}=\Psi_j(\ve x^{(i)}),\hspace{2mm} i=1\enum N;\hspace{1mm} j=1\enum \rm{card} \,\ca\}$. Then, the LOO error can be computed as
\begin{equation}
\label{eq:errLOO_simp}
Err_{LOO}=\frac{1}{N}\sum_{i=1}^N \left(\frac{\cm(\ve x^{(i)})-\cm^{\rm{PCE}}(\ve x^{(i)})}{1-h_i}\right)^2.
\end{equation}
The relative LOO error, $err_{LOO}$, is obtained by normalizing $Err_{LOO}$ with the empirical variance of $\cy=\{\cm({\ve x}^{(i)})\enum \cm({\ve x}^{(N)})\}$, denoting the set of model evaluations at the experimental design. Because this error estimate may be too optimistic, a corrected estimate is herein employed, given by \cite{Chapelle:2002}
\begin{equation}
\label{eq:errLOO_corr}
err^*_{LOO}=err_{LOO}\left(1-\frac{\rm{card}\ca}{N}\right)^{-1}\left(1+\rm{tr}((\ve \Psi^{\rm{T}}\ve \Psi)^{-1})\right).
\end{equation}
This corrected LOO error is a good compromise between fair error estimation and affordable computational cost.

\subsection{Sobol' indices from polynomical chaos expansions}

Let us consider the PCE $\widehat{Y}=\cm^{\rm{PCE}}(\ve X)$ of the quantity of interest $Y=\cm(\ve X)$. It is straightforward to obtain the Sobol' decomposition of $\widehat{Y}$ in an analytical form by observing that the summands $\cm_{\ve u}^{\rm{PCE}}(\ve X_{\ve u})$ in Eq.~(\ref{eq:Sobol_short}) can be written as
\begin{equation}
\label{eq:summands_PCE}
\cm_{\ve u}^{\rm{PCE}}(\ve X_{\ve u})=\sum_{\ua \in \ca_{\ve u}}{y_{\ua}}\ve {\Psi_{\ua}},
\end{equation}
where $\ca_{\ve u}$ denotes the set of multi-indices that depend \emph{only on} $\ve u$, \ie
\begin{equation}
\label{eq:summands_PCE_indices}
\ca_{\ve u}=\{\ua \in \ca: \alpha_k\neq 0 \hspace{1mm} \mathrm {if \hspace{1mm} and \hspace{1mm} only \hspace{1mm} if} \hspace{1mm} k \in \ve u\}.
\end{equation}
Clearly, $\bigcup \ca_{\ve u}=\ca$. Consequently, due to the orthonormality of the PCE basis, the partial variance $D_{\ve u}$ is estimated as
\begin{equation}
\label{eq:partial_var_PCE}
\widehat{D}_{\ve u}=\Var{\cm_{\ve u}^{\rm{PCE}}(\ve X_{\ve u})}=\sum_{\ua \in \ca_{\ve u}}{y_{\ua}}^2,
\end{equation}
and the total variance is estimated as
\begin{equation}
\label{eq:var_PCE}
\widehat{D}=\Var{\cm^{\rm{PCE}}(\ve X)}=\sum_ {\ua \in \ca \backslash \{\ve 0\}}{y_{\ua}}^2.
\end{equation}
Accordingly, the Sobol' indices of any order can be obtained by a mere combination of the squares of the PCE coefficients. For instance, the PCE-based first- and second-order Sobol' indices are respectively given by
\begin{equation}
\label{eq:Si_order1_PCE}
\widehat{S}_i=\sum_{\ua \in \ca_i}{y_{\ua}}^2/\widehat D,
\hspace{7mm}
\ca_i=\{\ua \in \ca: \alpha_i>0, \alpha_{j\neq i}=0\}
\end{equation}
and
\begin{equation}
\label{eq:Si_order12_PCE}
\widehat{S}_{ij}\sum_{\ua \in \ca_{ij}}{y_{\ua}}^2/\widehat D,
\hspace{7mm}
\ca_{ij}=\{\ua \in \ca: \alpha_i,\alpha_j>0, \alpha_{k\neq i,j}=0\},
\end{equation}
whereas the total Sobol' indices are given by
\begin{equation}
\label{eq:Si_tot_PCE}
\widehat{S}_i^T=\sum_{\ua \in \ca_i^T}{y_{\ua}}^2/\widehat D,
\hspace{7mm}
\ca_i^T=\{\ua \in \ca: \alpha_i>0\}.
\end{equation}

It is evident that once a PCE representation of $Y=\cm(\ve X)$ is available, the complete list of Sobol' indices can be obtained at a nearly costless post-processing of the PCE coefficients requiring only elementary mathematical operations.

\section{Results and discussion}
\label{sec:04}

\begin{figure}[!ht]
\begin{center}
\includegraphics[width=0.98\textwidth]{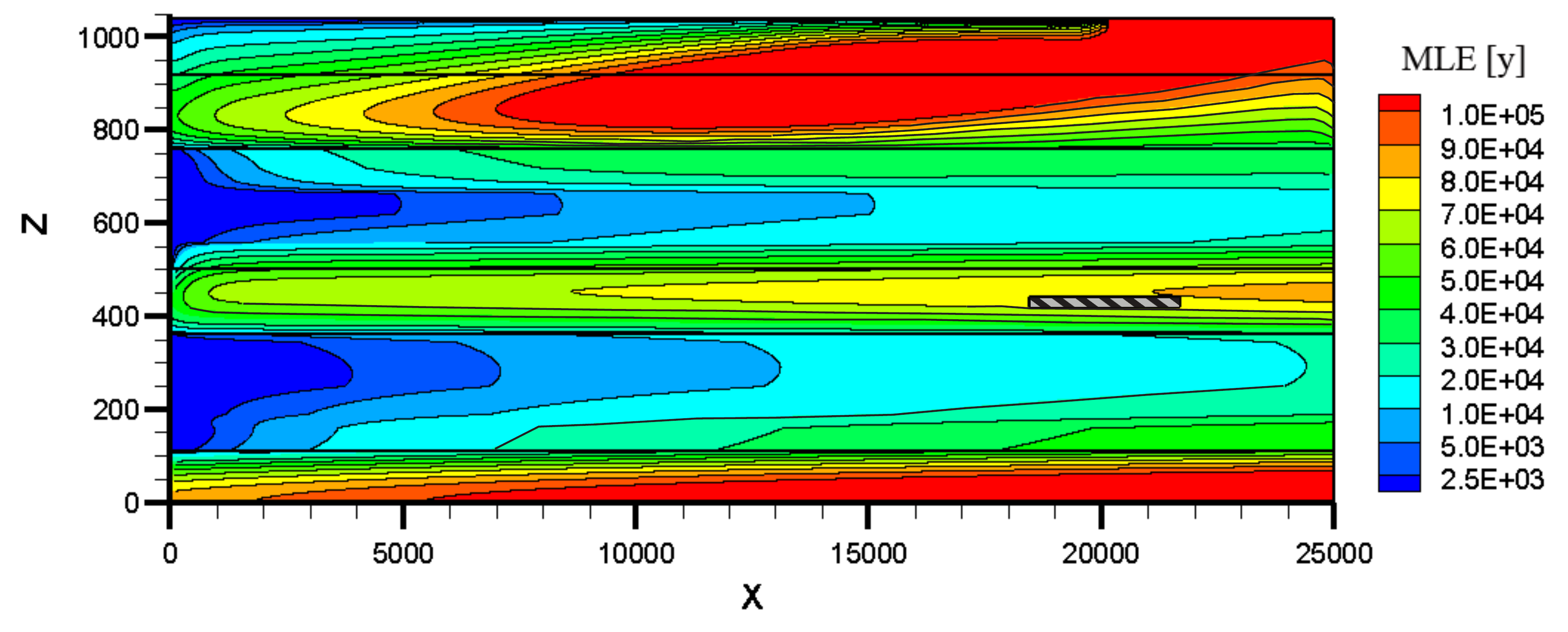}
\caption{Spatial distribution of mean lifetime expectancy in the reference case (vertical exaggeration: 10).}
\label{fig:NomMLE}
\end{center}
\end{figure}

Figure \ref{fig:NomMLE} provides an overview of the distribution of the MLE throughout the entire model domain in the nominal case. In this case, the parameters are distributed homogeneously in each of the 15~layers; $K_x$ and $\phi$ take on the values given in Table \ref{tab:NominalValues} for each layer, whereas for all layers the anisotropy ratio is $A_K = 0.1$, the Euler angle is $\theta = 0$ degree, the longitudinal component of the macro-dispersion tensor is $\alpha_L = 15$~m and the anisotropy ratio is $A_\alpha = 0.1$. The hydraulic gradients follow the boundary conditions settings described in Section \ref{sec:FlowBCs}.

Because of its highly confining properties, the middle layer (C2) presents values of MLE $>$ 40,000 years. On average, it takes approximately 75,000 years for a solute departing from the TZ to reach any  outlet of the domain. Much lower MLE values are found in the two aquifer sequences, with the Oxfordian displaying slightly smaller values. The effect of conductive layers is clearly distinguishable as fringes of low MLE values stretch in layers D4, L1a and L1b in particular. As a result of the low permeability in the top two layers (K3 and K1-K2) and the bottom layer (T), water molecules can take more than a 100,000 years to flow through the domain. 

In the following, we compute the PCE-based Sobol' indices for the MLE at the TZ by implementing the theory presented in Section \ref{sec:03}. We consider a high-dimensional random input encompassing the entirety of hydro-dispersive parameters described in Section \ref{sec:02} as well as the flow boundary conditions. Note that the term \emph{porosity}, $\phi$, is construed in the discussion of the Sobol' indices. Since the values of the hydraulic conductivities are retrieved through approximation functions, the estimation of the sensitivity for the $\phi$ variables is implicitly associated to that of the $K_x$ variables in the respective layers. This is singularly important when interpreting the Sobol' indices for aquifer formations, where the hydraulic conductivity governs the ageing process (Section \ref{sec:Discussion}). Computations of the PCE and Sobol' indices are performed with the uncertainty quantification software UQLab \cite{MarelliICVRAM2014, UQdoc_09_106}.

\subsection{Modeling of input uncertainty}
\label{sec:04-2}

The uncertain input in each of the 15 layers of the hydrogeological model comprises the following parameters governing the advective-dispersive processes: the petrofacies, $P$; the anisotropy in the components of the hydraulic conductivity tensor, $A_K$; the Euler angle of the hydraulic conductivity tensor, $\theta$; the longitudinal component of the dispersivity tensor, $\alpha_L$; and the anisotropy in the longitudinal and vertical components of the dispersivity tensor, $A_\alpha$. Furthermore, the uncertain input includes  the hydraulic gradients, $\nabla H$, at three zones: the Dogger sequence, the Oxfordian sequence and the top of the model domain, respectively denoted as zone~1, zone~2 and zone~3.

\begin{figure}[!ht]
\centering
\includegraphics[width=0.98\textwidth]{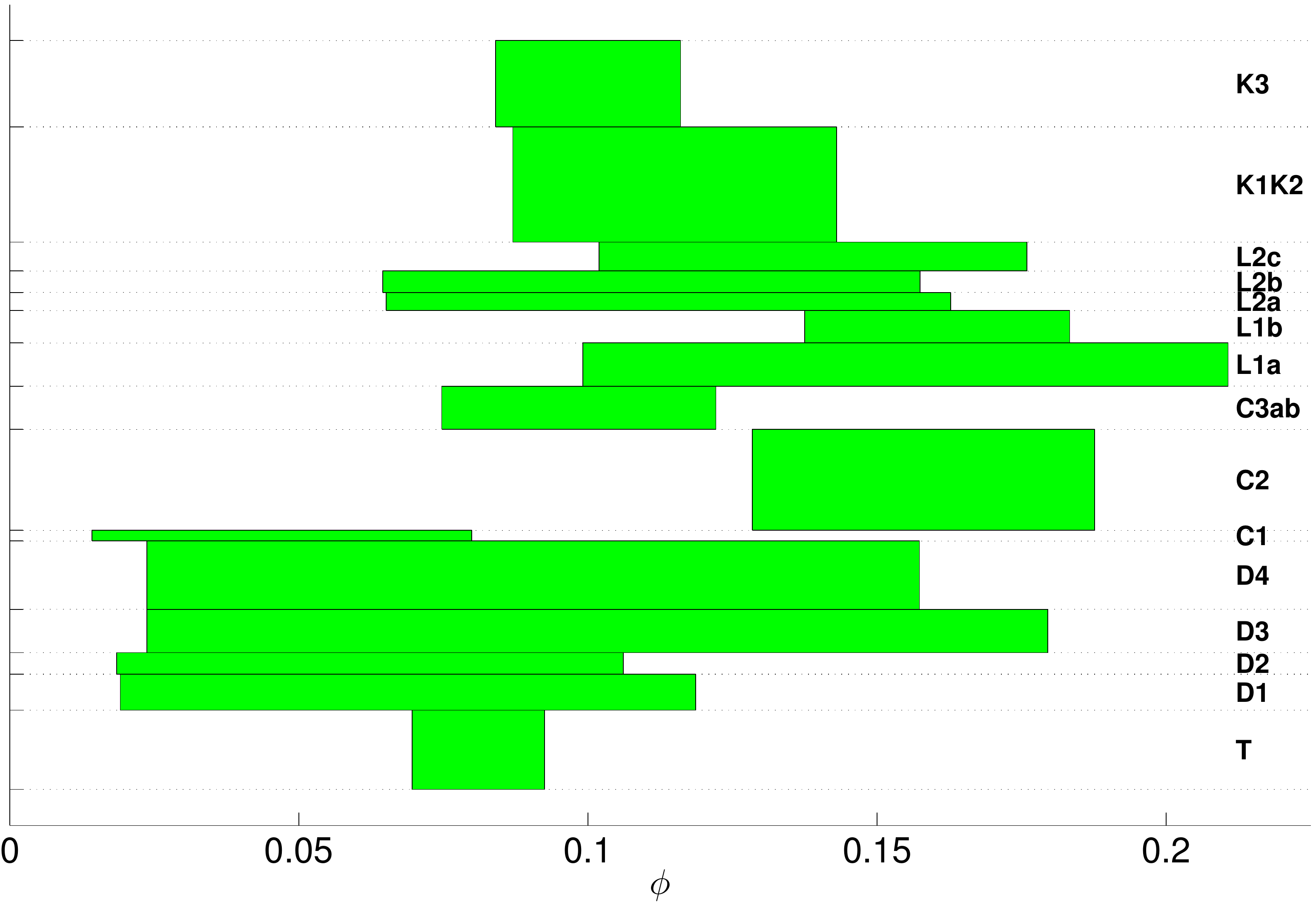}
\caption{Porosity ranges along the model cross-section.}
\label{fig:phi}
\end{figure}

\begin{figure}[!ht]
\centering
\includegraphics[width=0.98\textwidth]{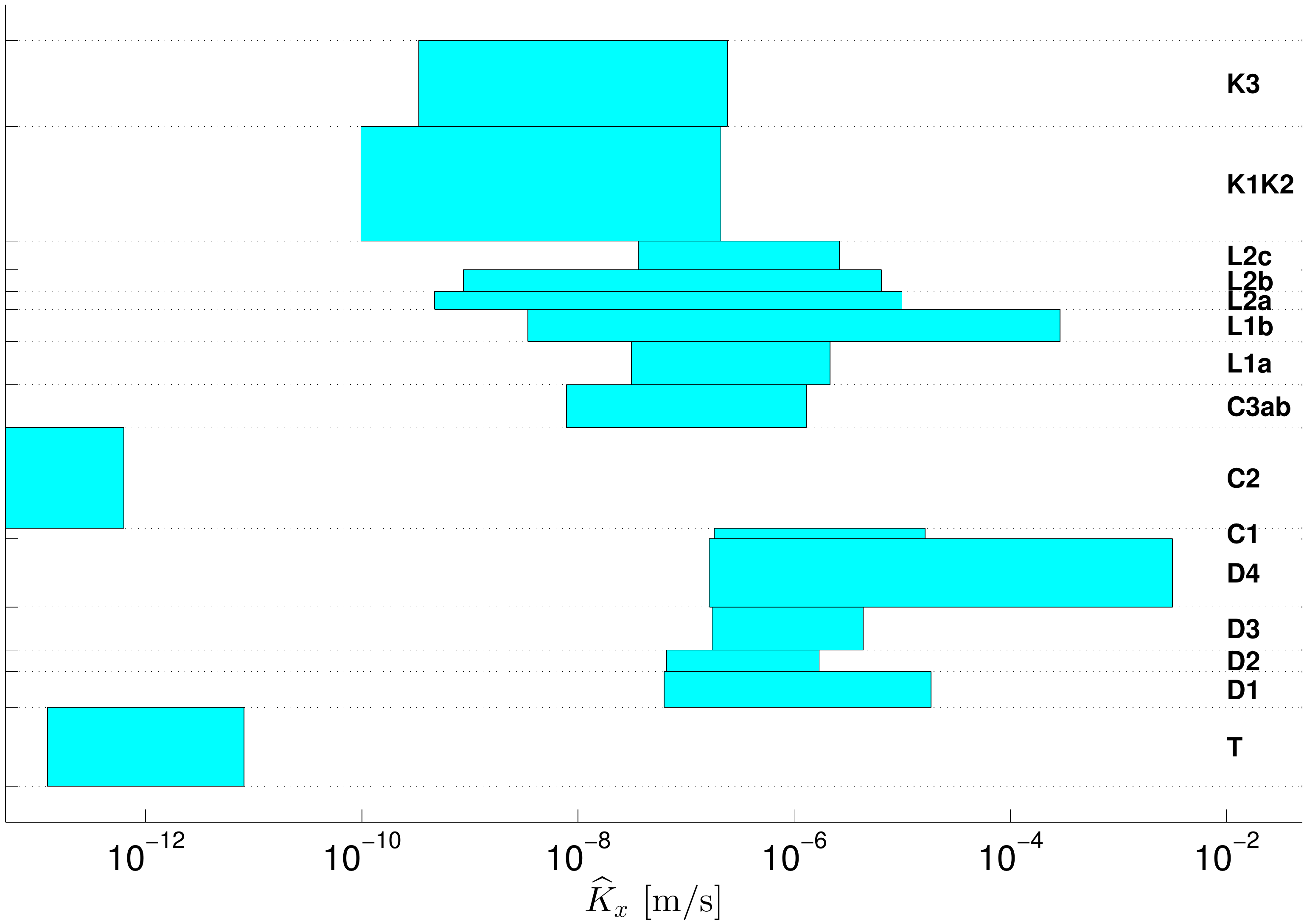}
\caption{Hydraulic conductivity ranges along the model cross-section.}
\label{fig:Kx}
\end{figure}

As stated before, a deterministic relationship $\log_{10}(K_x)=f(\phi)$ is assumed for each layer, \ie the uncertainty regarding the petrofacies, $P$, of a layer is treated through the porosity, $\phi$. 
The available \emph{sparse} data, characterizing specific locations in the subsurface of the Paris Basin, provide insufficient information for a comprehensive definition of the CDF of the $\phi$ parameters. Thus, in order to conduct a conservative study and avoid introducing any bias in the analysis, the uncertain porosities are modeled as uniform random variables. The porosity ranges are bounded by the values $\phi^{(min)}$ and $\phi^{(max)}$ listed in Table \ref{tab:UncertaintyRanges} together with the respective coefficients of variation (CoV); these are shown graphically along the model cross-section in Figure \ref{fig:phi}. The bounds $\phi^{(min)}$ and $\phi^{(max)}$ represent the 1st and 9th deciles of the corresponding CDF derived from porosity values measured in each geological layer. This approach is justified by the presence of local extreme measures that cannot be representative for the whole layer. Bounds for the $K_x$ parameters are also provided in Table \ref{tab:UncertaintyRanges}, consistently with the $\log_{10}(K_x)=f(\phi)$ approximation functions, and presented graphically in Figure \ref{fig:Kx}.

\begin{table}[!ht]
\centering
\caption{Ranges of porosity, $\phi$, and respective CoV and hydraulic conductivity values, $K_x$, in the 15 geological layers.}
\label{tab:UncertaintyRanges}
\begin{tabular}{c c c c c c}
\hline
Layer & $\phi^{(min)}$ [-] & $\phi^{(max)}$ [-] & CoV & $K_x^{(min)}$ [m/s] & $K_x^{(max)}$ [m/s]\\
\hline
K3 & 0.0840 & 0.1160 & 0.0924 & 3.3734e-10 & 2.4078e-07\\
K1-K2 & 0.0870 & 0.1430 & 0.1406 & 9.8116e-11 & 2.0928e-07\\
L2c & 0.1019 & 0.1759 & 0.1538 & 3.6186e-08 & 2.6212e-06\\
L2b & 0.0645 & 0.1574 & 0.2417 & 8.7318e-10 & 6.3950e-06\\
L2a & 0.0651 & 0.1627 & 0.2474 & 4.7005e-10 & 9.9336e-06\\
L1b & 0.1375 & 0.1833 & 0.0824 & 3.4324e-09 & 2.8913e-04\\
L1a & 0.0991 & 0.2107 & 0.2080 & 3.1165e-08 & 2.1523e-06\\
C3ab & 0.0747 & 0.1221 & 0.1391 & 7.8488e-09 & 1.2945e-06\\
C2 & 0.1284 & 0.1876 & 0.1082 & 5.0349e-14 & 6.2570e-13\\
C1 & 0.0142 & 0.0799 & 0.4031 & 1.8184e-07 & 1.6195e-05\\
D4 & 0.0237 & 0.1573 & 0.4262 & 1.6408e-07 & 3.1521e-03\\
D3 & 0.0237 & 0.1795 & 0.4427 & 1.7470e-07 & 4.3539e-06\\
D2 & 0.0185 & 0.1061 & 0.4059 & 6.6071e-08 & 1.7049e-06\\
D1 & 0.0191 & 0.1186 & 0.4172 & 6.2552e-08 & 1.8425e-05\\
T & 0.0696 & 0.0925 & 0.0816 & 1.2325e-13 & 8.1328e-12\\
\hline
\end{tabular}
\end{table}

Due to the lack of explicit information, each of the parameters $A_K$, $\theta$, $\alpha_L$ and $A_{\alpha}$ is identically distributed in the different layers. In particular, the anisotropy ratios $A_K$ and $A_\alpha$ both follow a uniform distribution in [0.01, 1], the Euler angle $\theta$ is uniformly distributed in [-30, 30] (in degrees) and the parameter $\alpha_L$ is uniformly distributed in [5, 25] (in meters). The static hydraulic gradients are also uniformly distributed in the ranges given in Table \ref{tab:UncertaintyGradH}. These were obtained by applying a perturbation of 20\% of the nominal hydraulic gradient within each of the three zones. Although hypothetical, these conservative uncertainty ranges were purposely chosen to provide insights into the behavior of the multi-layered model.

\begin{table} [!ht]
\centering
\caption{Ranges of hydraulic gradient at the three zones of interest.} \label{tab:UncertaintyGradH}
\begin{tabular}{c c c c}
\hline Zone number & $\nabla H^{(min)}$ & $\nabla H^{(max)}$ \\
\hline  1 & 0.00064 & 0.00096 \\
2 & 0.00240 & 0.00360 \\
3 & 0.00272 & 0.00408\\
\hline
\end{tabular}
\end{table}

In total, the model subsumes $M=78$ \emph{independent} input random variables upon the MLE of water molecules outflowing from the TZ in the middle of layer C2.

\subsection{Polynomial chaos expansions of the model response}

\begin{figure}[!ht]
\includegraphics[width=0.48\textwidth]{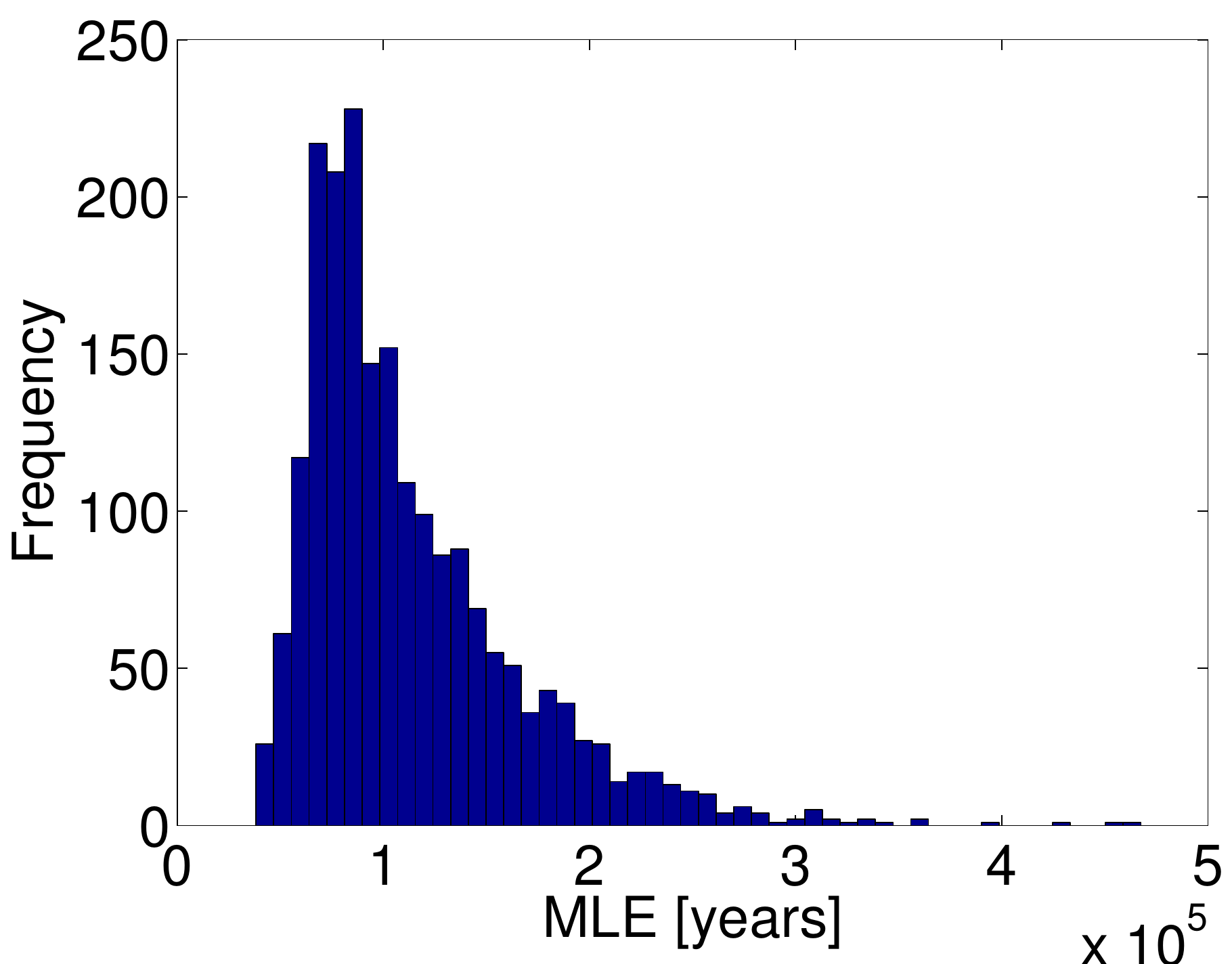}
\includegraphics[width=0.48\textwidth]{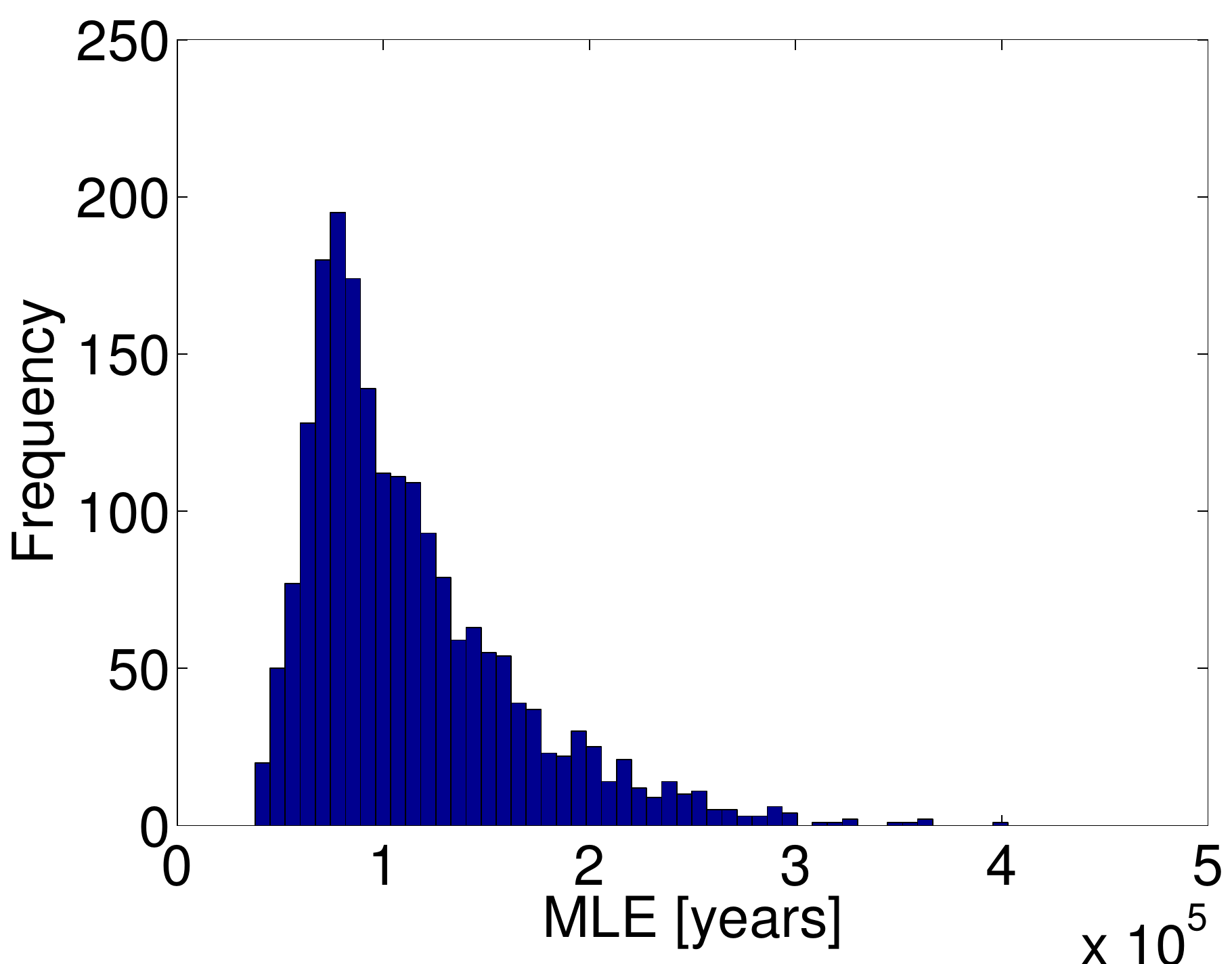}
\caption{Histograms of mean lifetime expectancy values calculated with sets $\mathbf{\ce}$ (left) and $\mathbf{\ce}'$ (right).}
\label{fig:MLEhist_2}
\end{figure}

In the sequel, we build PCE of the model response in terms of the $78$ input random variables. To this end, we employ Latin hypercube sampling (LHS) to draw the experimental design \cite{McKay1979}. LHS is a popular technique for obtaining random experimental designs ensuring uniformity of each sample in the domains of the margin input variables $\{X_1\enum X_M\}$. The data available herein for building the PCE consist of the MLE values for (i) an experimental design of size $N=2,000$, drawn with LHS and denoted $\ce$, and (ii) an enrichment of $\ce$ of equal size $N'=2,000$, denoted $\ce'$. The enrichment is built according to the nested-LHS approach so that the joint set $\{\ce,\ce'\}$ is approximately a LHS experimental design as well \cite{Wang2003, Blatman2010b}. Histograms of the model response for the two sets of input vectors are shown in Figure \ref{fig:MLEhist_2}. Positively skewed distributions are observed for both output sets with the modes situated at MLE $\approx$ 85,000 years.

We develop PCE based on $\ce$ and on the joint set $\{\ce,\ce'\}$  and assess their comparative accuracy. For the set $\ce$, we consider the MLE response in both the original and the logarithmic scales; in this case, the enrichment $\ce'$ serves as a validation set for computing the generalization error (see Section \ref{sec:err}). For all PCE, the candidate basis is determined using a hyperbolic truncation scheme with $q=0.5$ (see Eq.~(\ref{eq:trunc})). Sparse PCE are developed by varying the maximum degree $p$ from 1 to 15; the meta-model of optimal degree is selected as the one yielding the smallest corrected LOO error (see Eq.~(\ref{eq:errLOO_corr})).

\begin{figure}[!ht]
\centering
\includegraphics[width=0.48\textwidth]{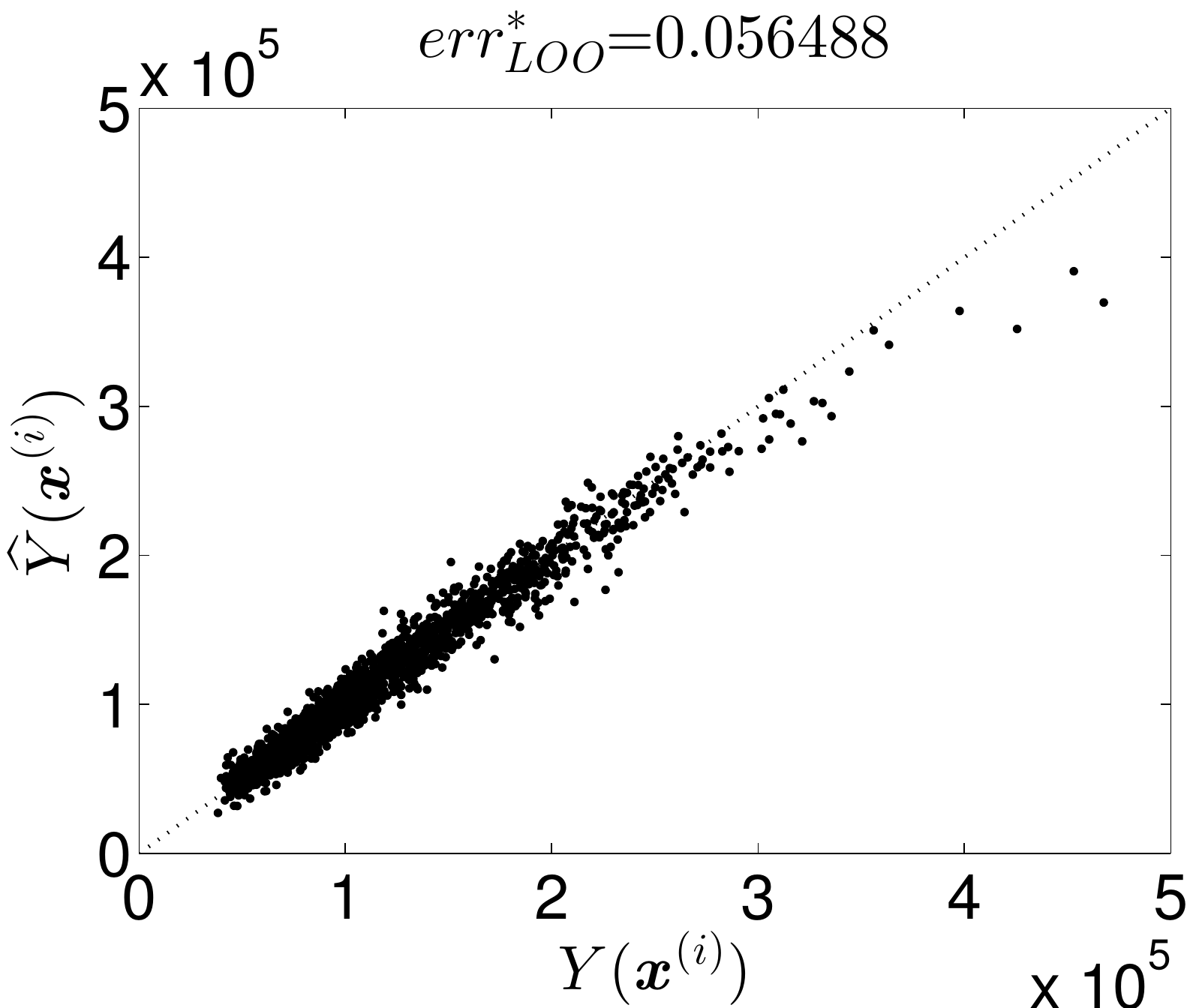}
\includegraphics[width=0.48\textwidth]{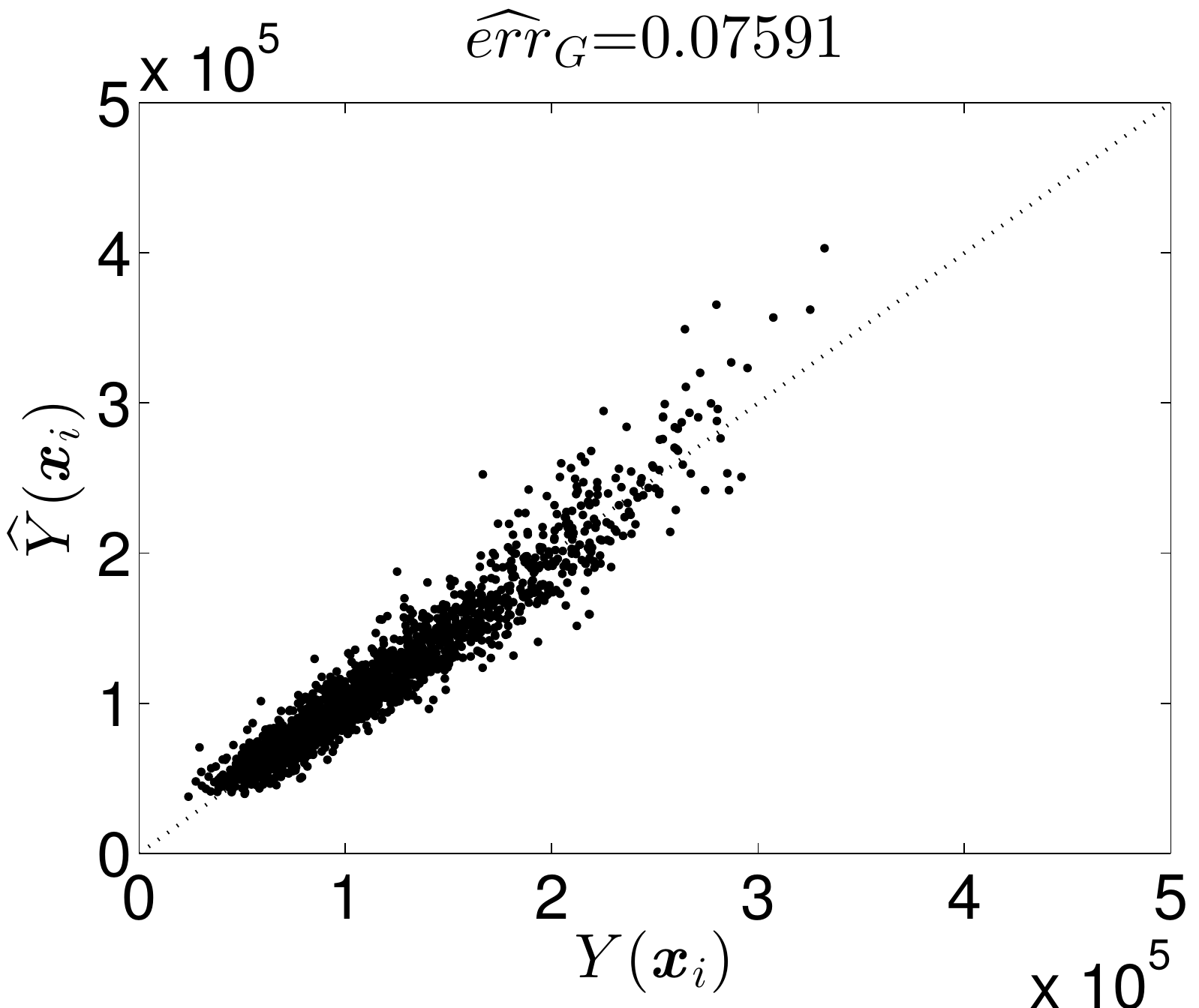}
\caption{Comparison of PCE A with the actual model response at the experimental design, $\ce$, (left) and at the validation set,  $\ce'$ (right).}
\label{fig:PCE_2A}
\end{figure}

The first PCE, denoted A, is built using $\ce $ as the experimental design and considering the MLE in the original scale. The optimal degree is $p=8$ and the corresponding corrected LOO error is $err^*_{LOO} = 0.0565$. The sparse PCE includes $185$ basis elements, whereas the total number of basis elements for $p=8$ and $q=0.5$ is $18,643$; for $q=1$, the size of the candidate basis would be $5.3\times10^{10}$.  The index of sparsity, defined as the number of basis elements in the sparse expansion divided by the size of a full basis for the same $p$ and $q$, is $185/18,643=9.9\times10^{-3}$. The small value of this index herein indicates the interest in developing sparse PCE for analyses of high-dimensional models. The sparse basis consists of polynomials in 68 out of the 78 total input parameters, meaning that the output does not depend at all on the values of the 10 excluded parameters. Note that 3 out of the 10 excluded parameters are properties of layer T. The estimate of the generalization error (evaluated with $\ce'$) is $\widehat{err}_G=0.0759$. The left graph of Figure \ref{fig:PCE_2A} compares the values of the meta-model, $\widehat{Y}$, with the respective values of the exact model, $Y$, at the input samples of the experimental design, $\ce$. A similar comparison but for the validation set, $\ce'$, is shown in the right graph of the same figure.

\begin{figure}[!ht]
\centering
\includegraphics[width=0.48\textwidth]{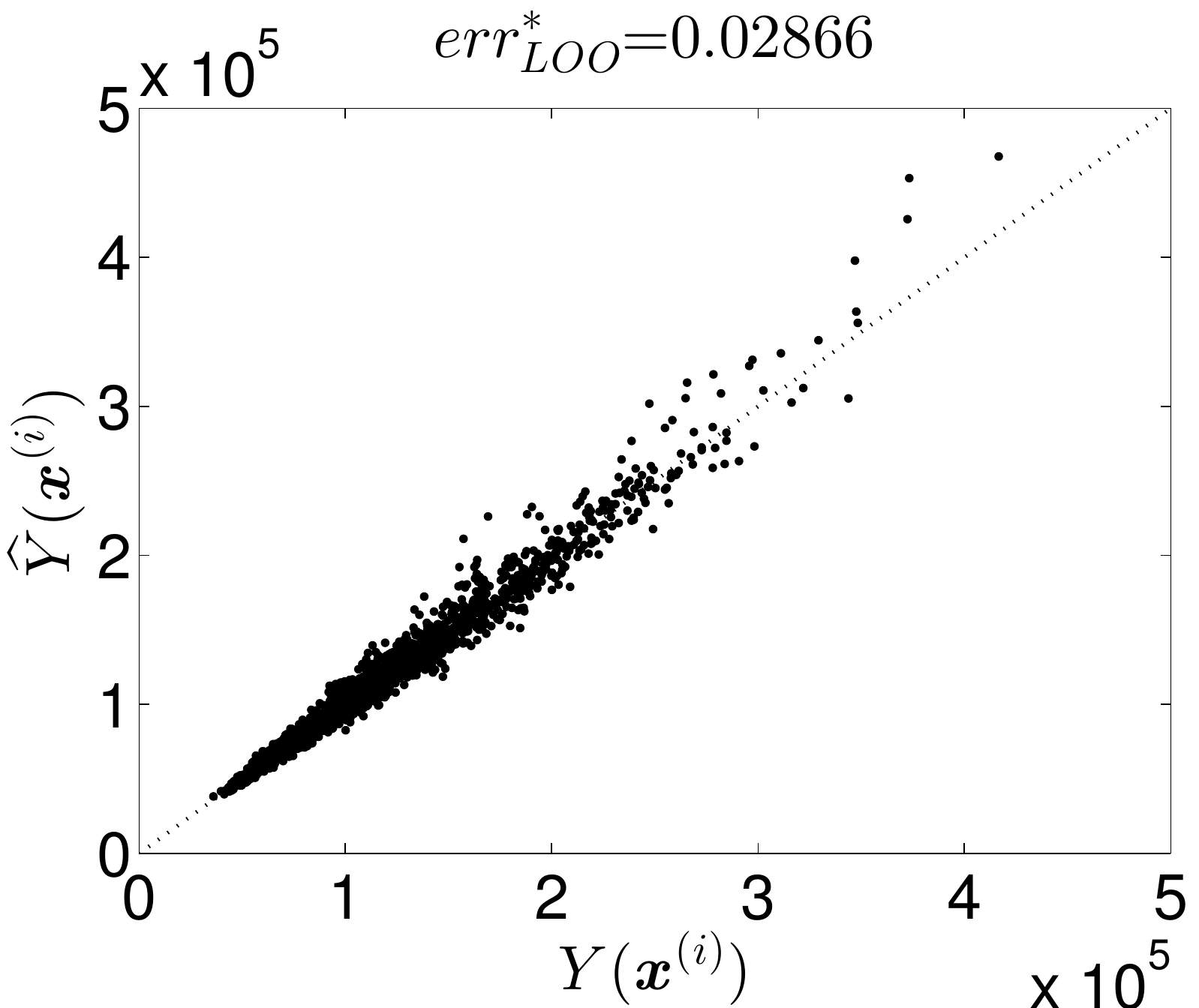}
\includegraphics[width=0.48\textwidth]{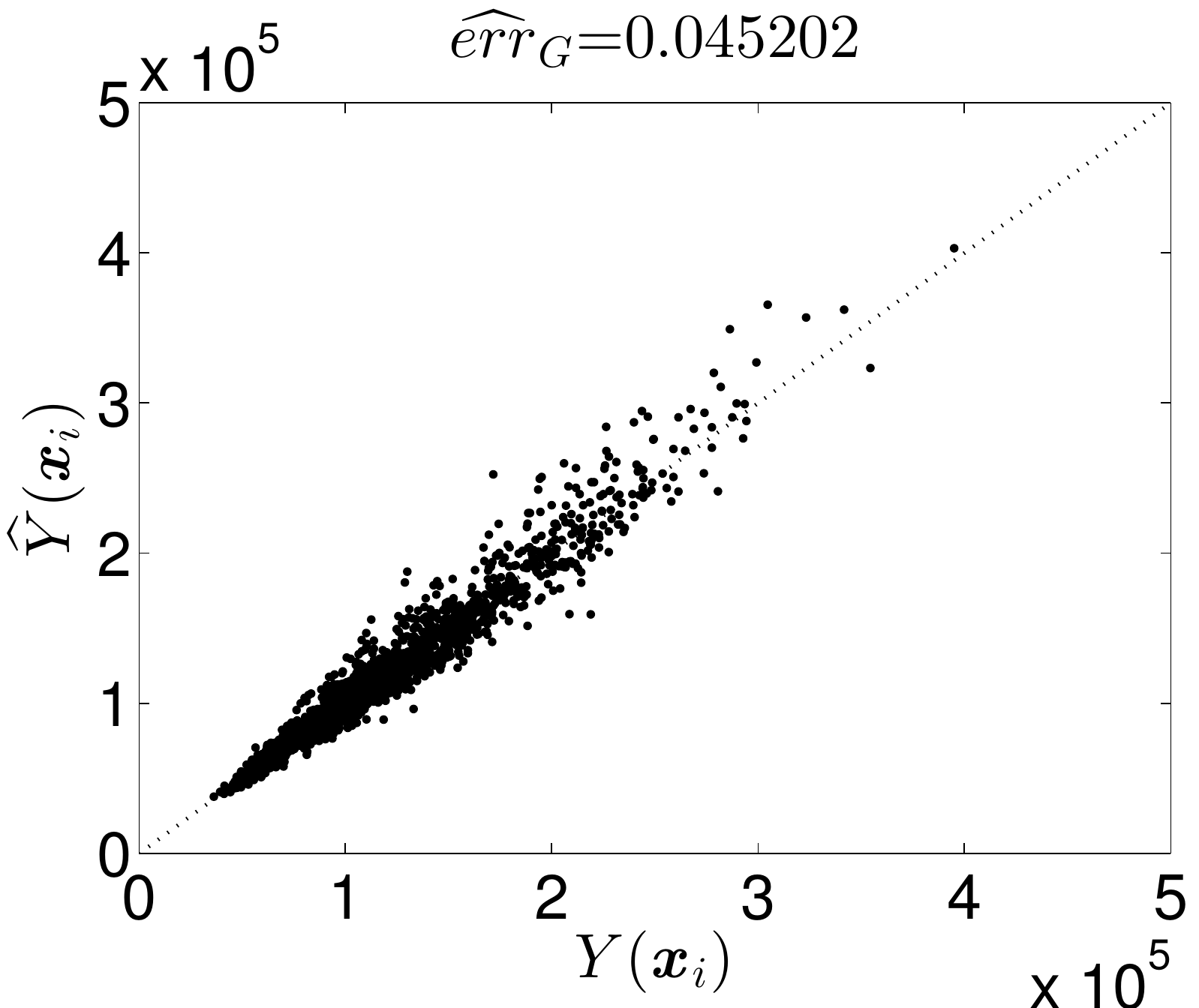}
\caption{Comparison of exponential transformation of PCE B with the actual model response at the experimental design, $\ce$, (left) and at the validation set, $\ce'$ (right).}
\label{fig:PCE_2B}
\end{figure}

A second PCE, denoted B, is built by using again $\ce$ as the experimental design, but employing a logarithmic transformation of the MLE. The optimal sparse PCE is obtained for $p=8$ (same degree as for PCE A) with corresponding corrected LOO error $err^*_{LOO} = 0.0287$. It comprises $163$ basis elements and thus, has an index of sparsity $163/18,643=8.7\times10^{-3}$. The sparse basis consists of polynomials in 65 out of the 78 total input parameters, with 6 out of the 13 excluded parameters representing dispersivity anisotropy ratios ($A_\alpha$). The resulting generalization error for the MLE response in the original scale (considering the exponential transformation of the obtained PCE) is estimated as $\widehat{err}_G=0.0452$. Note that both $err^*_{LOO}$ and $\widehat{err}_G$ are lower than the respective error estimates for PCE A. The left and right graphs of Figure \ref{fig:PCE_2B} compare the exponential transformation of PCE B with the exact model response at $\ce$ and $\ce'$, respectively.

\begin{figure}[!ht]
\centering
\includegraphics[width=0.48\textwidth]{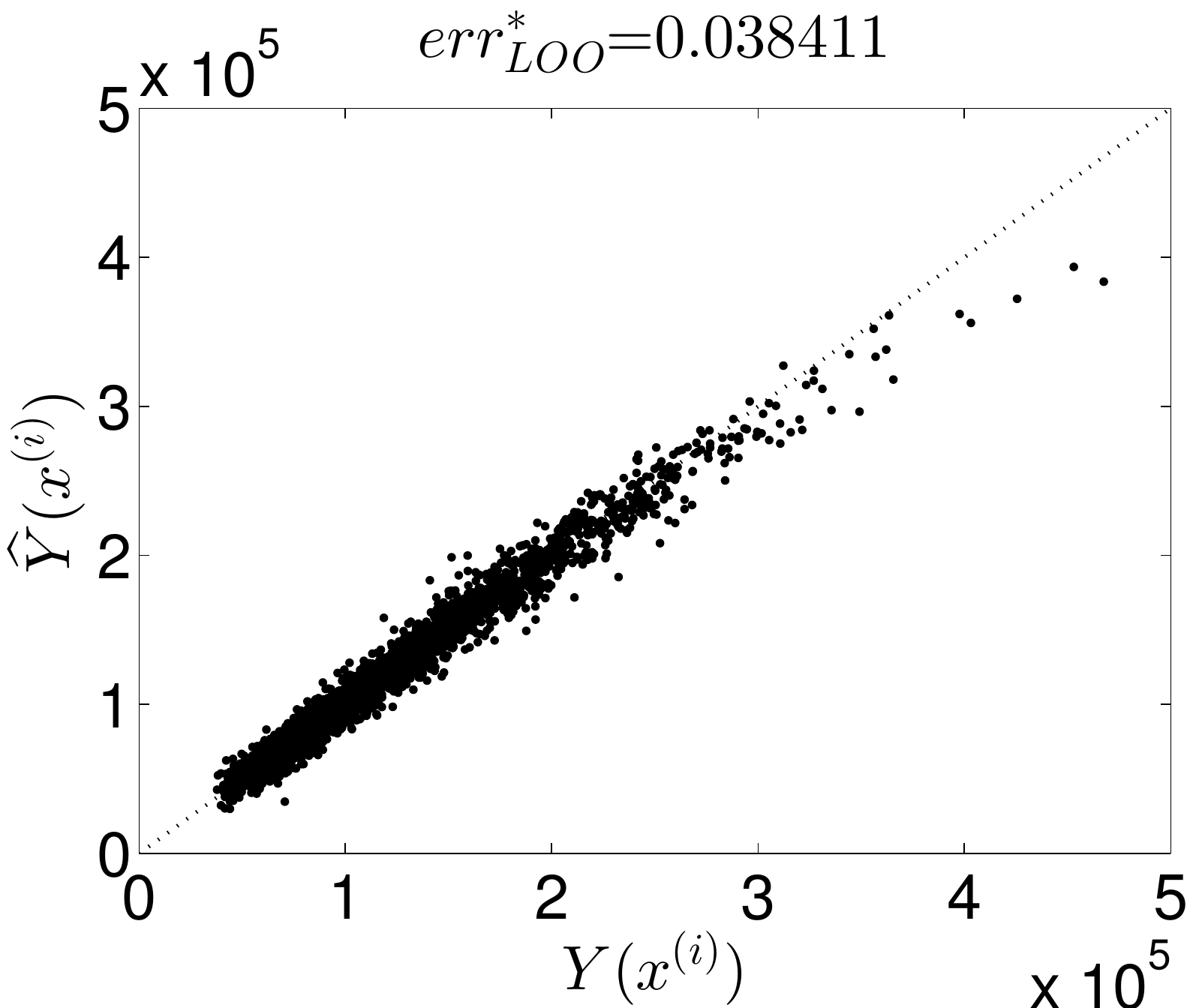}
\caption{Comparison of PCE C with the actual model response at the experimental design, $\{\ce,\ce'\}$.}
\label{fig:PCE_2C}
\end{figure}

Finally, we use as experimental design the joint set $\{\ce,\ce'\}$ consisting of $N+N'=4,000$ points. The optimal PCE is obtained for $p=10$ and the corresponding corrected LOO error is $err^*_{LOO} =  0.0384$. The sparse PCE comprises $312$ basis elements and has an index of sparsity $312/106,887\approx 2.9\times 10^{-3}$. The only two parameters excluded from the sparse basis are $A_K^{L2b}$ and $\alpha_L^T$. The comparison between this meta-model, denoted C, and the exact model response at the input samples of the experimental design, $\{\ce,\ce'\}$, is shown in Figure \ref{fig:PCE_2C}.

Assessing the relative accuracies of the three meta-models, we note that all have (corrected) LOO errors of the same order of magnitude, with the smallest error corresponding to PCE B. Because it is of interest to limit the number of costly evaluations of the exact hydrogeological model, an experimental design comprising $2,000$ points is deemed most appropriate. We therefore conduct SA for the MLE response by post-processing the coefficients of PCE A and B and compare the results. We underline that the Sobol' indices are not invariant under a logarithmic transformation of the response quantity of interest (see \eg \cite{borgonovo2014}). Nevertheless, we expect that the Sobol' indices for the non-transformed and the logarithmic MLE will exhibit similar trends, \ie a parameter that is important for the variance of the MLE will also be important for the variance of its logarithm. Moreover, it is of interest to perform SA with PCE B, because the logarithmic transformation represents herein a meaningful quantity from a physical viewpoint considering the wide variation of MLE. 

To further elaborate on the accuracy of the two meta-models employed in the subsequent SA, we compare the ED-based mean, $\mu$, and standard deviation, $\sigma$, of the response with the respective quantities, $\widehat{\mu}$ and $\widehat{\sigma}$, obtained from the PCE coefficients. For the response in the original scale (PCE A), we have $\{\mu=113,216;~\sigma=53,220\}$ and $\{\widehat{\mu}=113,410;~\widehat{\sigma}= 50,735\}$, corresponding to errors $\{\epsilon_{\widehat{\mu}}= 0.17~\%;~ \epsilon_{\widehat{\sigma}}= -4.67~\%\}$. For the logarithmic response (PCE B), we have $\{\mu=11.545;~\sigma=0.4180
\}$ and $\{\widehat{\mu}=11.543;~\widehat{\sigma}= 0.4073\}$, leading to errors $\{\epsilon_{\widehat{\mu}}= -0.01
~\%;~\epsilon_{\widehat{\sigma}}= -2.56~\%\}$. Overall, PCE B is more accurate.

\subsection{Sensitivity analysis}

\begin{figure}[!ht]
\centering
\includegraphics[width=0.9\textwidth]{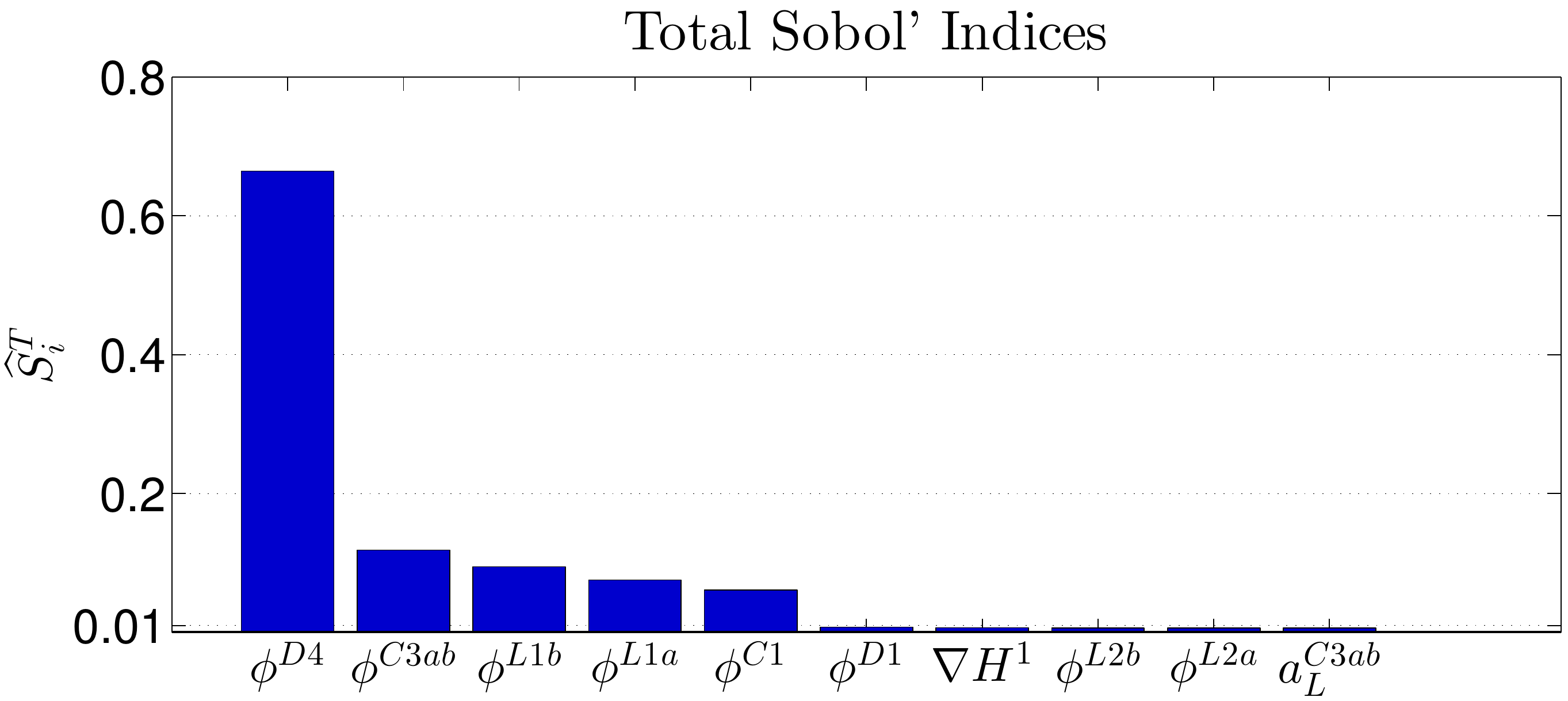}
\\
\vspace{2mm}
\includegraphics[width=0.9\textwidth]{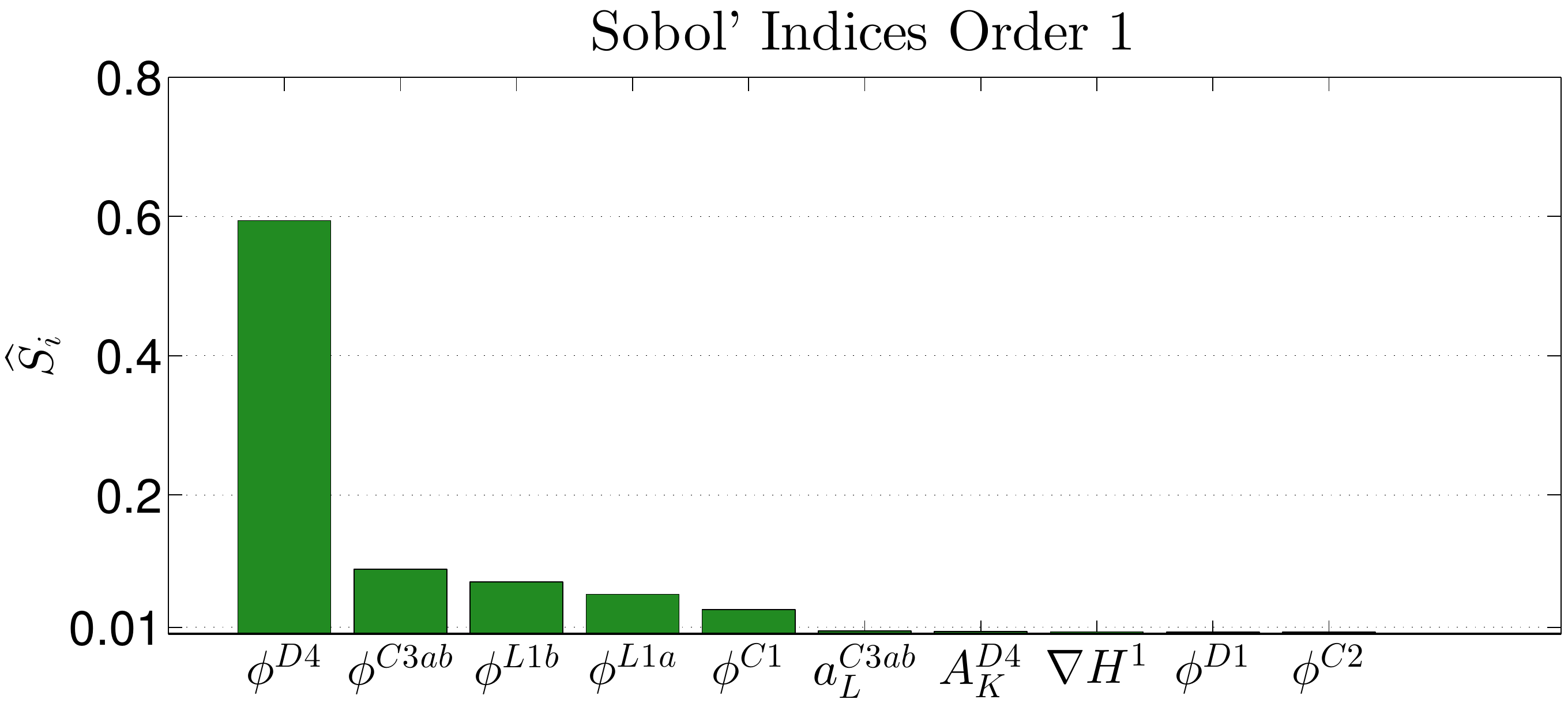}
\caption{Sobol' indices using PCE A.}
\label{fig:Sobol_2A}
\end{figure}

Figure \ref{fig:Sobol_2A} shows bar-plots of the total and first-order Sobol' indices for PCE A. The ten largest indices are presented in descending order. The superscripts on the parameter symbols on the horizontal axes denote layer names or zone numbers. The figures indicate the same ranking of the five most important parameters in terms of both the total and first-order indices.  These are the porosities of layers D4, C3ab, L1b, L1a and C1 in order of importance, with the porosity of layer D4 being dominant. Employing the criterion $S^T_i<0.01$ to sort out unimportant variables, the porosities of the aforementioned five layers comprise the only important parameters. Therefore, the screening allows one to consider 73 out of 78 parameters as unimportant, meaning that they could be given a deterministic value without affecting essentially the predicted MLE.

We note that all five layers with porosities identified as important are located close to the host layer C2. D4 is the thickest among those and has the highest hydraulic conductivity. Although C1 is adjacent to the host layer C2, it is associated with smaller total and first-order indices than other neighboring layers, which may be attributed to its small thickness.

The largest second-order effects (not shown) comprise interactions between the five porosities classified above as important and involve one of the layers D4 or L1b. The second-order effects explain $10.9\%$ of the total variance, whereas contributions from higher-order effects are practically zero.

\begin{figure}[!ht]
\centering
\includegraphics[width=0.9\textwidth]{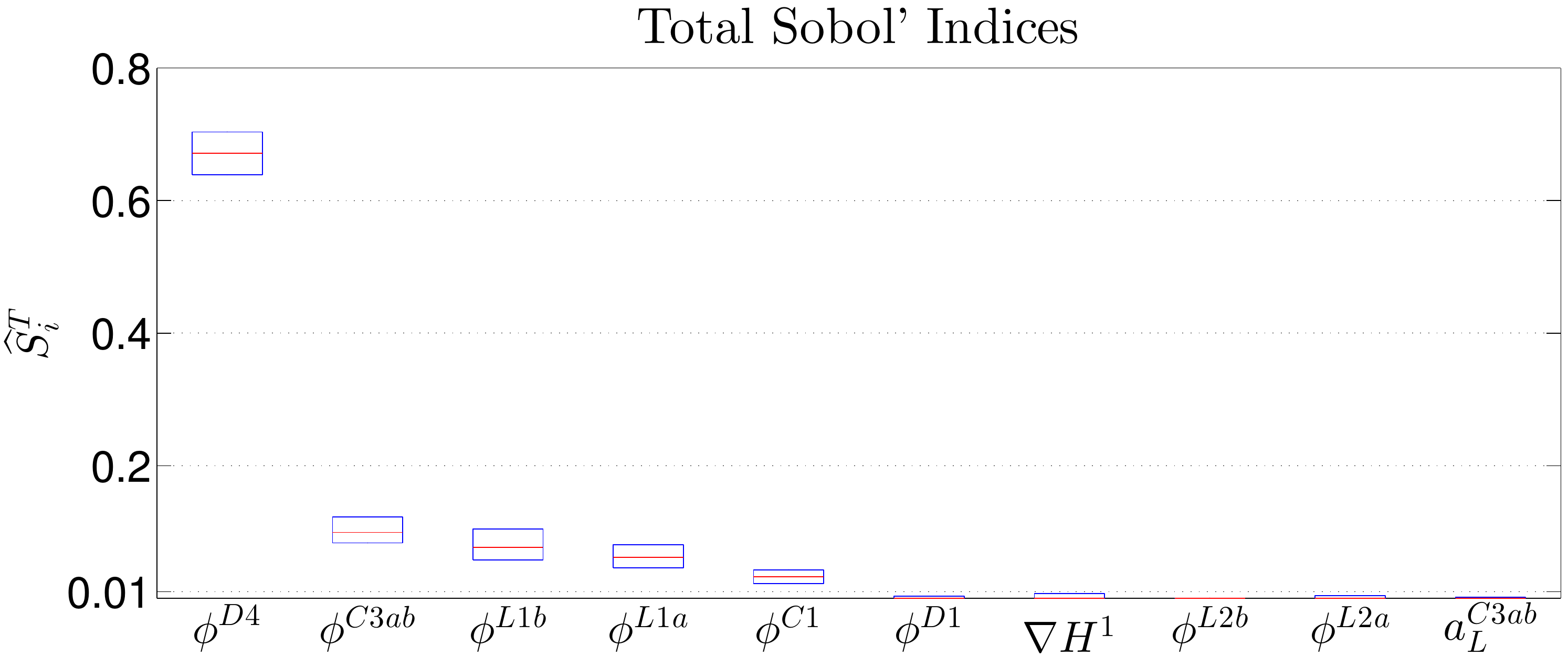}
\caption{Boxplot of total Sobol' indices for the MLE in the original scale using experimental designs with 200 points.}
\label{fig:Boxplot}
\end{figure}

In the sequel, we demonstrate that the sensitivity pattern in this high-dimensional problem can be detected by only a few runs of the model. To this end, we randomly extract 200 points out of the experimental design $\ce$ and follow the same procedure as above to compute the PCE-based sensitivity indices for the MLE in the original scale. Conducting 100 repetitions, we obtain statistics for the ten largest total Sobol' indices identified in the analysis with PCE A (see upper graph of Figure \ref{fig:Sobol_2A}). These are depicted in the boxplot in Figure \ref{fig:Boxplot}, in which the central mark of a box indicates the median and its edges indicate the 25th and 75th percentiles. We observe that the median Sobol' indices from SA performed with experimental designs comprising \emph{only} 200 points are fairly close to the Sobol' indices obtained with the experimental design $\ce$ comprising 2,000 points. Moreover, the dispersion is relatively small, \ie $\pm 0.03$ at most, which is more than sufficient for a preliminary screening of the important variables. These results emphasize the robustness and remarkable efficiency of our approach in dealing with high-dimensional SA problems.

\begin{figure}[!ht]
\centering
\includegraphics[width=0.9\textwidth]{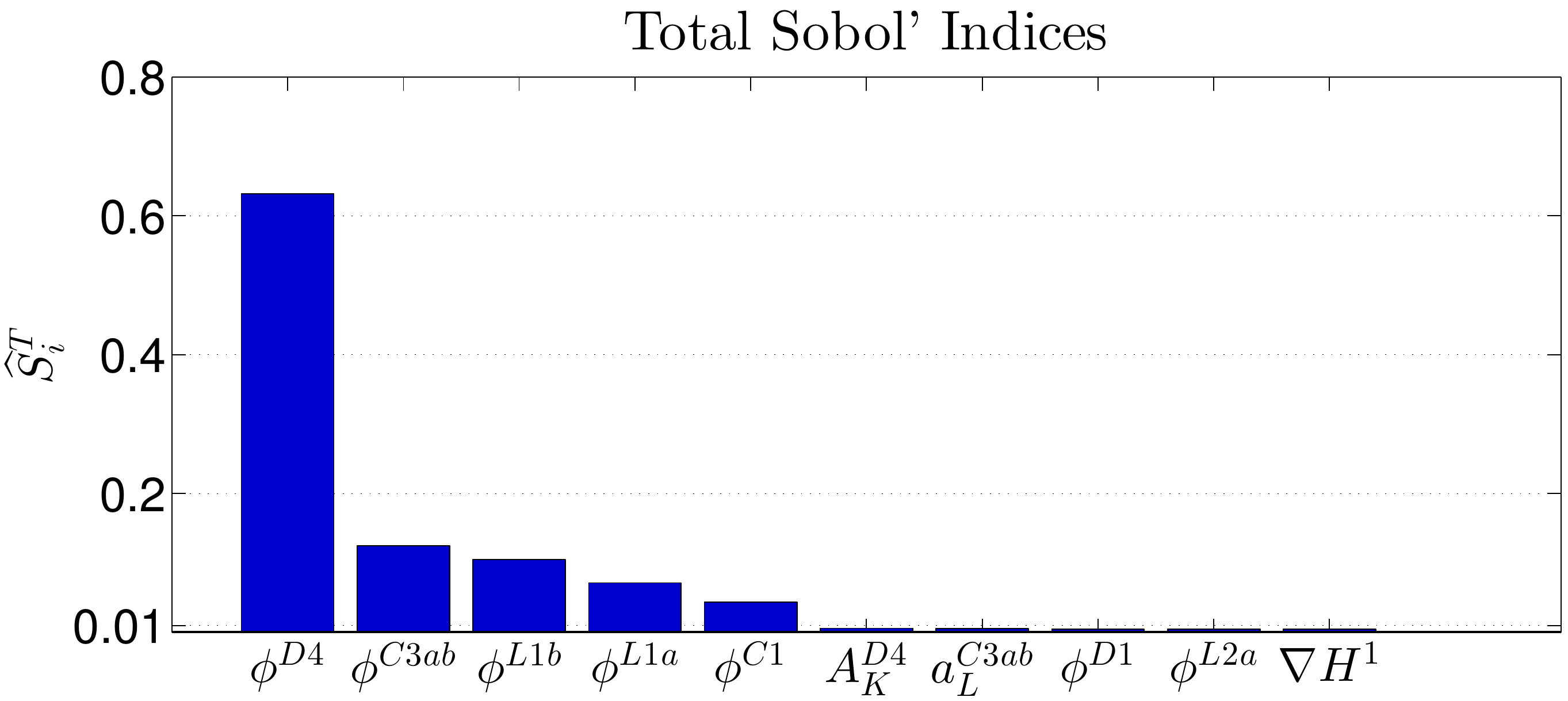}
\\
\vspace{2mm}
\includegraphics[width=0.9\textwidth]{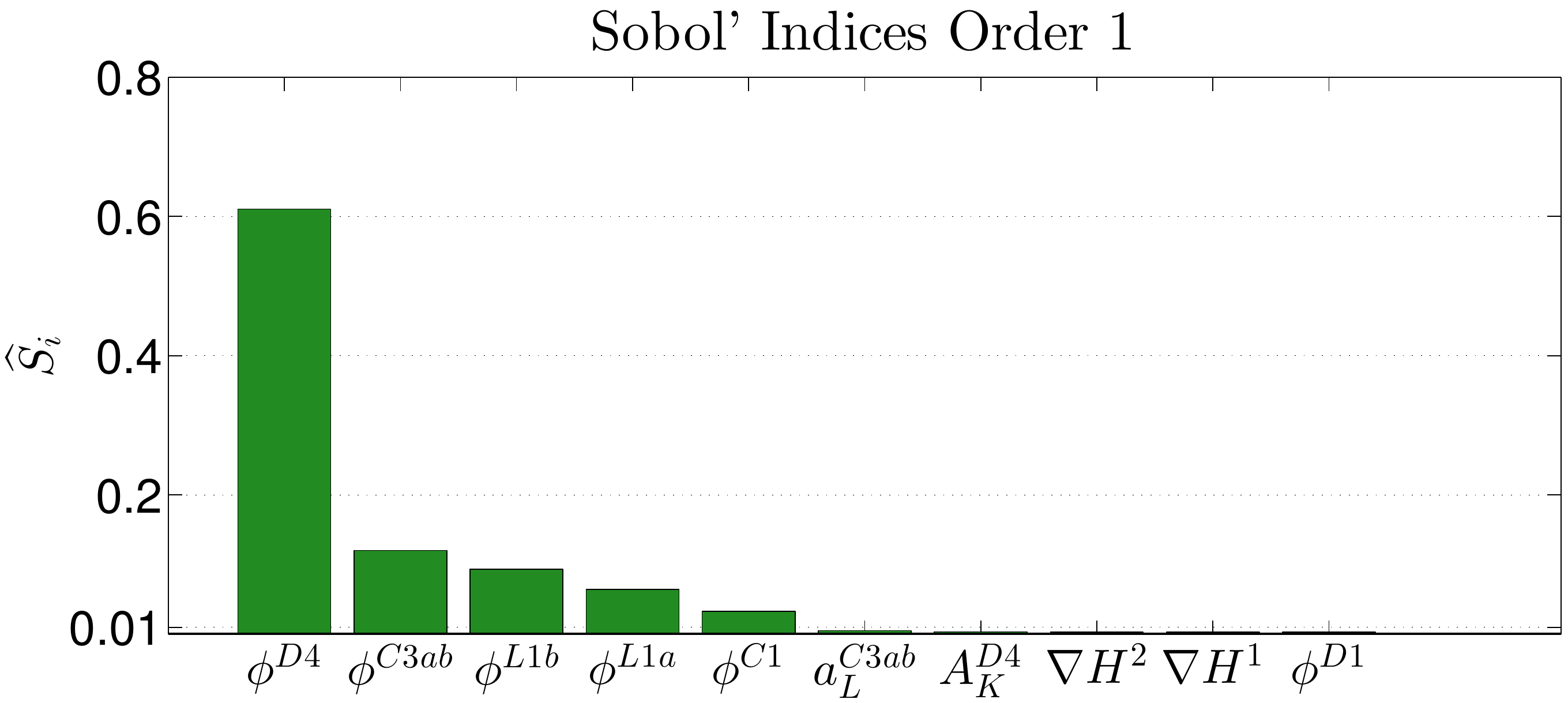}
\caption{Sobol' indices using PCE B.}
\label{fig:Sobol_2B}
\end{figure}

Let us now compare the above results with respective results obtained by considering the logarithmic MLE. In Figure \ref{fig:Sobol_2B}, we show bar-plots of the ten highest total and first-order Sobol' indices in descending order for PCE B. Obviously, the indices obtained in the two cases, \ie  considering the non-transformed and the logarithmic MLE, follow similar trends. SA with PCE B identifies the same five variables as important ($S^T_i\geq0.01$) in the same order as the analysis with PCE A, with the porosity of layer D4 being dominant. Furthermore, the five important variables are also the ones with the highest first-order indices following the same ranking. Again, the largest second-order effects involve the porosities of one of the layers D4 or L1b, while their sum explains $4.2\%$ of the total variance. Thus, second-order effects are slightly less significant for the logarithmic response than for the response in the original scale. Contributions of higher than second-order effects are practically zero.

\begin{figure}[!ht]
\centering
\includegraphics[width=0.95\textwidth]{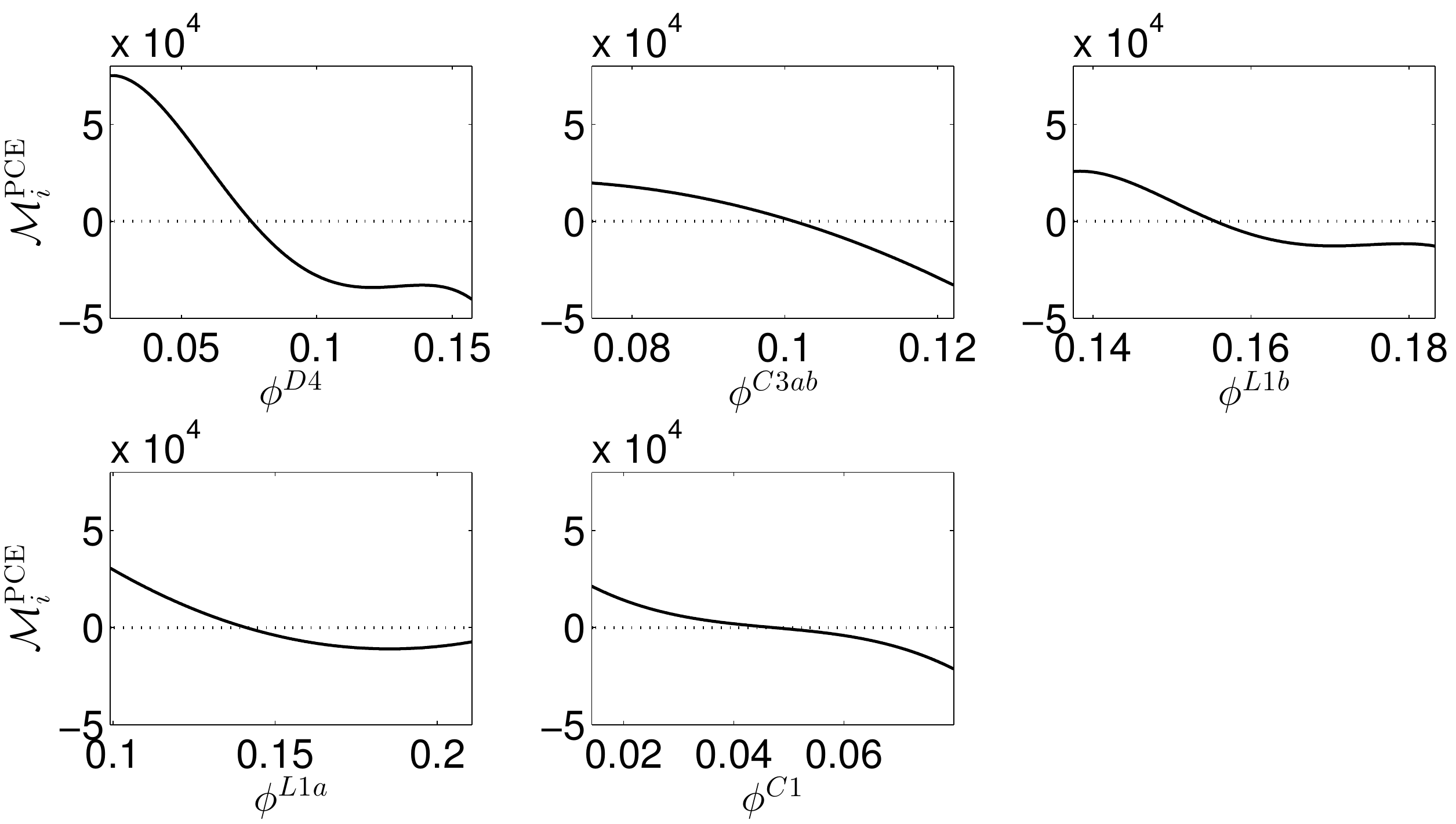}
\caption{Univariate effects of important variables using PCE A.}
\label{fig:UniEff_2A}
\end{figure}

\begin{figure}[!ht]
\centering
\includegraphics[width=0.95\textwidth]{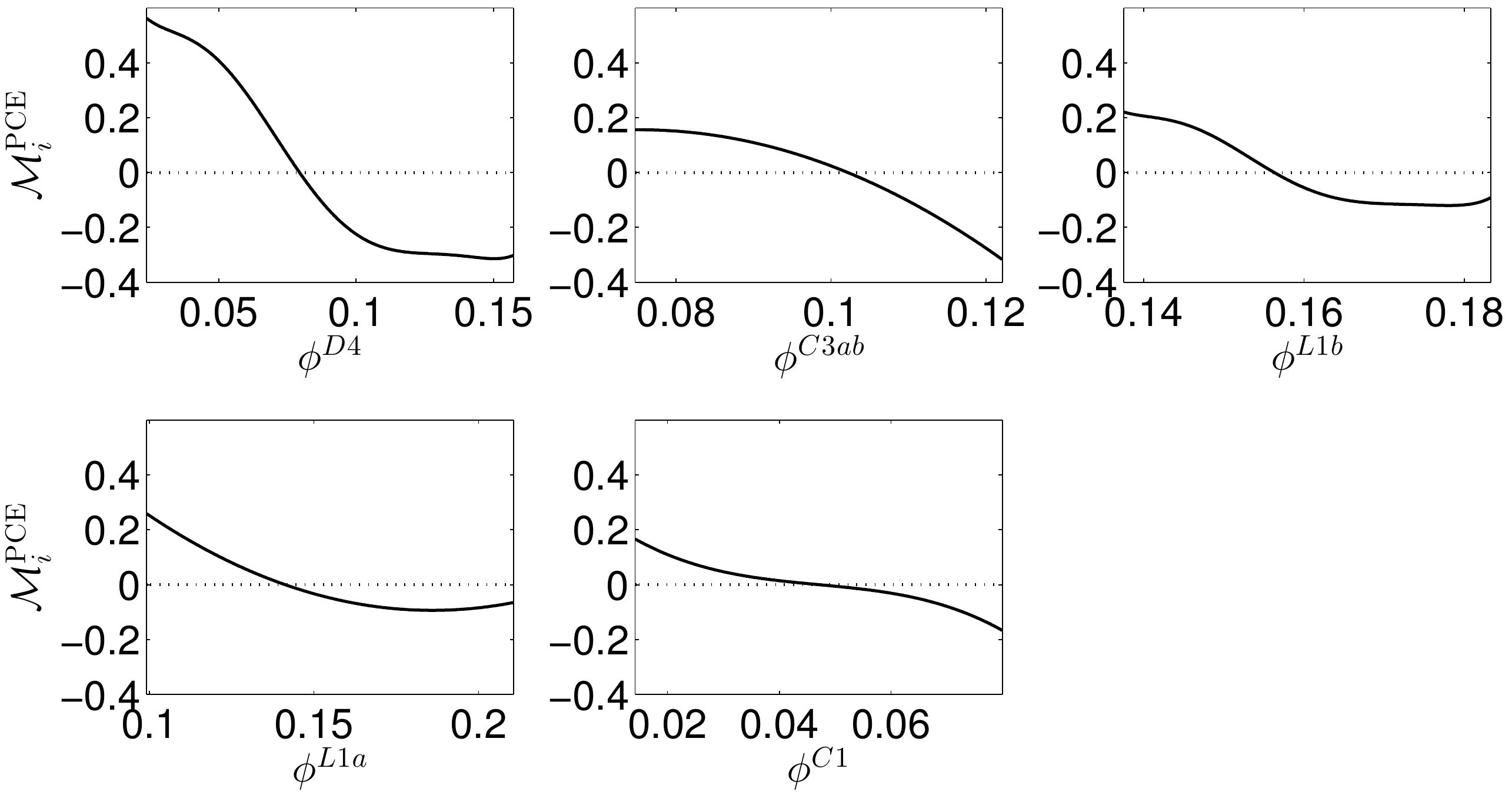}
\caption{Univariate effects of important variables using PCE B.}
\label{fig:UniEff_2B}
\end{figure}

To gain further insight into the effects of the important variables on the model response, we examine the behavior of the corresponding first-order summands comprising univariate polynomials only, \ie
\begin{equation}
\label{eq:UniEff_exp}
\cm_i^{\rm PCE}(x_i)= \Espe{}{\cm^{\rm PCE}(\ve X|X_i=x_i)}
\end{equation}
or equivalently
\begin{equation}
\label{eq:UniEff}
\cm_i^{\rm PCE}(x_i) = \sum_{\ua \in \ca_i}{y_{\ua}} \Psi_{\ua}(x_i),\hspace{7mm}
\ca_i=\{\ua \in \ca: \alpha_i>0, \alpha_{i\neq j}=0\}.
\end{equation}
Figures \ref{fig:UniEff_2A} and \ref{fig:UniEff_2B} depict such univariate effects considering PCE A and B, respectively, for the porosities of the five layers classified as important. Despite the different scales, the shapes of respective curves in the two figures demonstrate similar trends. The two PCE include univariate polynomials of the same degree in $\phi^{C3ab}$, $\phi^{L1a}$ and $\phi^{C1}$ (up to second or third degree), but PCE B includes higher-degree univariate polynomials in $\phi^{D4}$ and $\phi^{L1b}$ (up to sixth degree) than PCE A (up to fourth degree). The figures demonstrate that an increasing porosity and thus, an increasing hydraulic conductivity are associated with a decreasing algebraic contribution to the MLE value. Indeed, the hydraulic conductivity parameter, mainly responsible for the advective processes within the layer, is linearly and oppositely related to the MLE (see Eq.~(\ref{eq:Flux}) and (\ref{eq:LE})). Overall, the response is more sensitive to changes occurring within low porosity-permeability ranges, where advective processes are counterbalanced by dispersive-diffusive processes, with the latter being responsible for higher MLE values. Plateaus observed in certain graphs indicate the ascendancy of one ageing process over the other, which results in more regular trends.

For a more in-depth investigation of the contributions of the different types of hydro-dispersive parameters, Table \ref{tab:SumsIndices} lists the sums of the first-order indices per type of property over all layers or zones of the considered cross-section. According to this table, the added main effects of the porosity parameters account for approximately $87\%$ and $93\%$ of the response variance for PCE A and B, respectively, whereas the added main effects of all remaining input random variables account for a mere $2.5\%$ and $2.8\%$, respectively. We note the higher contributions of main effects for PCE B and the zero main effects of $A_\alpha$ for both PCE.

\begin{table} [!ht]
\centering
\caption{Sums of first-order Sobol' indices over all layers per type of property.} \label{tab:SumsIndices}
\begin{tabular}{c c c c c c c}
\hline  meta-model & $\phi$ & $A_K$ & $\theta$ & $\alpha_L$ & $A_\alpha$ & $\nabla H$\\
\hline  A & 0.8664 & 0.0088 & 0.0029 & 0.0076 & 0      & 0.0057\\
B & 0.9302 & 0.0096 & 0.0032 & 0.0088 & 0      & 0.0061\\
\hline
\end{tabular}
\end{table}

The above analysis indicated that among the set of random variables, only the porosity parameters of certain layers are important for explaining the response variance. As highlighted earlier, because of the assumed relationship between porosity and hydraulic conductivity in each layer, the Sobol' indices for the $\phi$ variables are also indicative of the importance of $K_x$ in the respective layers. Accordingly, in the subsequent discussion of the SA results, we will interpret the variability of MLE in terms of joint effects of the petrofacies $P$.

\subsection{Discussion of results}
\label{sec:Discussion}

The petrofacies of aquifer formations have the largest effects on the variability of the MLE at the TZ. For layers D4 and L1b in particular, for which the upper bounds of the hydraulic conductivity ranges are the highest (see Table \ref{tab:UncertaintyRanges}, Figure \ref{fig:Kx}), strong advective processes within their own volume may reduce the overall groundwater residence time in the model. Besides, the wider the ranges of $K_x$ values in these permeable formations, the higher are the contributions of the respective petrofacies to the variability of the MLE. The three most significant aquifer formations, namely D4, L1b and L1a, have rather non-linear univariate effects on the output response (see Figures \ref{fig:UniEff_2A} and \ref{fig:UniEff_2B}). Substantial changes within the response are observed with small shifts in the ranges of low porosity-permeability values, revealing the effect of the balance between advective and dispersive-diffusive transport processes. But for higher porosity-permeability values, advective fluxes prevail, thus yielding more linear and moderate changes in the response.

The position of the aquifers relatively to layer C2 is also relevant: the further the layer is, the lower is its effect on the MLE of water molecules departing from the TZ. Layer L1a, which is the first aquifer encountered in the Oxfordian sequence (at a distance of 60 meters from layer C2), has a significant effect on the output variance, whereas layers D2 and D3 that have similar $K_x$ ranges have much smaller contributions.

The relatively large uncertainty range and high upper-bound value of permeability for layer D1 results in a marginal effect on the response variability (see Figures \ref{fig:Sobol_2A} and \ref{fig:Sobol_2B}), which is nevertheless unimportant with respect to the threshold $S^T_i<0.01$. Its remote location relative to layer C2 explains the reduced quantity of water molecules departing from the TZ that reach the layer, given that most of the solute is captured by the highly advective aquifer from Upper Bathonian (D4) situated in between.

The petrofacies of semi-permeable formations, $P^{C3ab}$ and $P^{C1}$, are also significantly influencing the variability of the MLE at the TZ. By isolating layer C2 from major aquifer formations, they buffer solute intrusions into the Oxfordian and Dogger aquifer sequences, thus acting like a geological barrier. Figures \ref{fig:UniEff_2A} and \ref{fig:UniEff_2B} show that their univariate effects on the output quantity are relatively non-linear despite their limited amplitude.

We underline that the petrofacies of the host layer attributed to the Callovo-Oxfordian claystone ($P^{C2}$) are insignificant with regard to the variability of the ouput response. This feature was already outlined in the SA performed upon the integrated high-definition model in \citep{Deman2015}. Slow diffusive processes take place into highly impermeable rocks, which induces large values of the MLE response (Figure \ref{fig:NomMLE}) and prove the high efficiency of the Callovo-Oxfordian claystone to act as a geological barrier. Modifying the magnitudes of advective-dispersive transport processes in layer C2 does not add a significant variability to the time required for solutes to leave the domain where the numerical model is defined.

The magnitude of the transverse advective fluxes in each layer is related to the respective value of $K_z$, the uncertainty of which is accounted for through the anisotropy ratio $A_K$. Although  $A_K$ represents the second most influential group of factors considering added effects from all layers (see Table \ref{tab:SumsIndices}), the uncertainty in this property adds a small amount of variability to the MLE ($< 1\%$) compared to the petrofacies ($\approx 90\%$). We note however that factor $A_K^{D4}$ is only marginally excluded from the important factors when PCE B is considered (see Figure \ref{fig:Sobol_2B}). Indeed, for the highest $K_x$ values, strong advective fluxes take place within the layer's volume. Under the assumption of strong transverse fluxes ($A_K^{D4} \to 1$), solutes can be oriented toward neighbouring layers where slower fluxes occur, thus raising the MLE.

For each layer, the Euler angle $\theta$ could deviate groundwater fluxes from an orientation parallel to the $x$-axis and toward the main discharge boundaries, thus raising the variability of the response. Although it could be assumed as especially influential in the most advective layers, the total contribution of this group of uncertain factors to the variance of the MLE is negligible in comparison to that of the petrofacies $P$ (see Table \ref{tab:SumsIndices}).

In aquifer formations, the effects of the uncertainty regarding the macro-dispersion tensors upon the response quantity, \ie the magnitude ($\alpha_L$) and anisotropy ($A_\alpha$) of the tensors, are concealed by the strong effect of petrofacies on the advective part of the transport processes (see Table \ref{tab:SumsIndices}). We note however that the longitudinal component of the macro-dispersion tensor in layer C3ab ($\alpha_L^{C3ab}$) appears among the ten factors with the highest total Sobol' indices for both PCE A and B. The anisotropy ratios, $A_\alpha$, have no contribution at all to the response variance when considered independently; the uncertainty in these factors contributes to the variability of MLE only through interaction terms.

The sensitivity of the MLE with respect to flow BCs considered in the model is directly related to the magnitude and orientation of the advective fluxes in the entire model. In the case of high gradients in both limestone sequences, the advective solute transport processes would raise within their volume, and thus reduce the MLE. Table \ref{tab:SumsIndices} indicates that the three random hydraulic gradients have a small added effect on the output variance. Note however that the total Sobol' index for the hydraulic gradient in the Dogger sequence ($\nabla H^1$) belongs to the ten highest indices for both PCE A and B. The uncertainty regarding this factor can alter the advective processes occurring notably in layer D4, which has the far highest contribution to the variance of the MLE calculated at the TZ.

\section{Conclusions}
\label{sec:05}

The model introduced in this paper stands as a two-dimensional vertical cross-section of the subsurface of Paris Basin in the vicinity of Bure (Haute-Marne). While encompassing most of the hydrogeological features of the underground media, it was simplified with regard to geometries, discontinuities, fractures and heterogeneities. This numerical model is intended to explore the behavior of a complex multi-layered hydrogeological system at low computational cost and provide insights into the effect of uncertain parameters upon the mean lifetime expectancy (MLE).

Sensitivity analysis (SA) was carried out considering a high-dimensional random input. For the sake of simplicity, homogeneous parameters were assumed within each of the 15 hydrogeological layers comprising the model. The uncertain factors at each layer included: the petrofacies, $P$, regarded as the couple permeability-porosity, $\{K,\phi\}$; the anisotropy ratio and the orientation of the hydraulic conductivity tensor, $A_K$ and $\theta$ respectively; the magnitude and anisotropy ratio in the macro-dispersion tensor, $\alpha_L$ and $A_\alpha$ respectively. Additionally, the hydraulic gradients, $\nabla H$, in three zones of the model domain were considered random, leading to a total of 78 uncertain input factors.

In the present study, a target zone (TZ) located within the middle layer (C2) of the model domain provides the output response of interest. Latin hypercube sampling was employed to address the propagation of the uncertainty from the input factors upon the MLE of water molecules departing from the TZ. Polynomial chaos expansion (PCE) meta-models were used to compute the Sobol' sensitivity indices for each input factor at low computational costs. Sparse PCE proved singularly efficient in providing accurate representations of the response of interest at low computational cost with respect to the high dimensionality of the model. The accuracy was enhanced when the PCE were fitted to the logarithmic MLE; because of the wide range of variation of the MLE, considering its logarithmic transformation is herein meaningful from a physical standpoint.

The SA results for the non-transformed and the logarithmic MLE demonstrated similar trends. It was found that the variability of the MLE is almost entirely due to the uncertainty regarding the petrofacies of the hydrogeological layers. The other hydro-dispersive parameters are insignificant for explaining the response variance and may be considered as deterministic factors in future works. 

Focusing on the effects of petrofacies solely, the SA demonstrated the large contributions of aquifer formations to the variance of the model output. In particular, (i) the closer the aquifer formation to layer C2, (ii) the thicker the layer, (iii) the wider the ranges of the petrofacies and (iv) the higher the upper bound of the hydraulic conductivity range, the larger were the effects of the petrofacies on the variability of the response. Investigation of the univariate effects of petrofacies highlighted that for these permeable formations, the response is more sensitive to changes occurring within low porosity-permeability ranges. Hence, within a certain range of $\{K,\phi\}$ values, the dispersive-diffusive processes counterbalance with the strong advective fluxes in the ageing process.

In formations characterized by highly advective processes, the longitudinal hydraulic conductivities applying in the main groundwater direction have large contributions to the MLE variability. The two semi-confining formations encompassing the C2 layer buffer the transfer of solute from the latter toward the further aquifer sequences. Besides, it is acknowledged that longitudinal dispersion processes occurring within their own volume also retard the solute transfer toward the adjacent aquifers. Because of the diffusion-dominated transport processes occurring within its volume, the petrofacies of the highly-confining C2 layer have a negligible effect on the variance of the output response although responsible for the high values of the latter.

It is important to remind that the use of a 2D model tends to underestimate the output response of interest by omitting the advection and dispersion along the third dimension. Recognizing this limitation, we underline that the purpose of the simplified model introduced herein is to shed light on the relative effects of various uncertain factors governing the advective and dispersive processes in a complex multi-layered hydrogeological system. The presented methodology may be applied to a real-case study employing a realistic 3D numerical model. 

The sensitivity analysis performed in this work is deemed particularly informative for future applications with the high-resolution integrated Meuse/Haute-Marne hydrogeological model. In the frame of a real-case uncertainty analysis with concern to a solute transport in the subsurface of the Paris Basin, the authors recommend defining as thoroughly as possible the spatial distributions of hydraulic conductivity values, with a main focus on the large aquifer sequences of Oxfordian and Dogger ages closer to the potential host layer.\\

\textbf{Acknowledgement}

The authors are grateful to Mr Benjamin Brigaud (Universit\`{e} Paris-Sud) and Mrs Agn\`{e}s Vinsot (ANDRA) for providing the hydro-dispersive datasets used in this analysis, and to Mr Fabien Cornaton (DHI-WASY GmbH) for providing the code \emph{GroundWater}.

\bibliographystyle{model3-num-names}
\bibliography{biblio}

\medskip
\medskip

\label{lastpage}

\end{document}